\documentclass[10pt,letterpaper,aps,prl,twocolumn,superscriptaddress,floatfix]{revtex4-1}
\usepackage[latin1]{inputenc}
\usepackage{amsmath}
\usepackage{amsfonts}
\usepackage{amssymb}
\usepackage{bm}
\usepackage[pdftex]{graphicx}
\usepackage[colorinlistoftodos]{todonotes}  
\usepackage{graphicx}
\usepackage{subfigure} 

\usepackage{varioref}
\usepackage{braket}
\usepackage{esint}
\usepackage{mathtools}
\newcommand{\ra}{\rangle}
\newcommand{\be}{\begin{equation}}
\newcommand{\ee}{\end{equation}}

\begin{document}
\title{Quantum parameter estimation with the Landau-Zener transition}

\author{Jing Yang}
\affiliation{Department of Mechanical Engineering, University of Rochester, Rochester, NY 14627, USA}
\author{Shengshi Pang}
\affiliation{Department of Physics and Astronomy, University of Rochester, Rochester, NY 14627, USA}
\affiliation{Center for Coherence and Quantum Optics, University of Rochester, Rochester, NY 14627, USA}

\author{Andrew N. Jordan}
\affiliation{Department of Physics and Astronomy, University of Rochester, Rochester, NY 14627, USA}
\affiliation{Center for Coherence and Quantum Optics, University of Rochester, Rochester, NY 14627, USA}
\affiliation{Institute for Quantum Studies, Chapman University, 1 University Drive, Orange, CA 92866, USA}

\date{\today}
\begin{abstract}
We investigate the fundamental limits in precision allowed by quantum mechanics from Landau-Zener transitions, concerning Hamiltonian parameters.  While the  Landau-Zener transition probabilities depend sensitively on the system parameters, much more precision may be obtained using the acquired phase, quantified by the quantum Fisher information.  This information scales with a power of the elapsed time for the quantum case, whereas it is time-independent if the transition probabilities alone are used.  We add coherent control to the system, and increase the permitted maximum precision in this time-dependent quantum system.  The case of multiple passes before measurement, ``Landau-Zener-Stueckelberg interferometry'', is considered, and we demonstrate that proper quantum control can cause the quantum Fisher information about the oscillation frequency to scale as $T^4$, where $T$ is the elapsed time.  
\end{abstract}
\maketitle

The Landau-Zener (LZ) transition is a classic example of exactly solvable, time-dependent quantum mechanics, whereby an effective two-level quantum system prepared in its ground state may either stay in the ground state, or transition to the excited state, depending on the speed of the energy separation of the levels \cite{landau_quantum_1981,zener_non-adiabatic_1932,stueckelberg1932theory,majorana1932atomi}. LZ transitions have been extended to parabolic level crossing \cite{suominen_parabolic_1992}, finite time duration with various approximation regimes \cite{vitanov_landau-zener_1996}, multi-level transitions such as those encountered in cavity and circuit QED \cite{sun2016landau,PhysRevA.93.063859}, and have also been studied in the presence of noise \cite{kayanuma1985stochastic,PhysRevB.67.134403,PhysRevB.76.024310}.
In the context of quantum information, the LZ transition has been used as a qubit readout mechanism and for quantum control \cite{PhysRevLett.94.057004,petta2010coherent,quintana2013cavity}.

The LZ transition has been used as a way of estimating Hamiltonian parameters, such as the level splitting energy, or the speed of the transition through the avoided level crossing \cite{wernsdorfer2000landau,PhysRevB.87.195412,izmalkov2004observation}.   Going beyond the LZ transition probabilities, it is also possible to make multiple, coherent sweeps of the avoided level crossing to accumulate a phase, also known as Landau-Zener-Stueckelberg interferometry \cite{shytov2003landau,oliver2005mach,shevchenko2010landau,sun2010tunable,PhysRevLett.96.187002,gaudreau2012coherent}; the acquired phase depends sensitively on the system parameters.  The field of quantum metrology is concerned with the optimal precision quantum physics permits in the estimation of parameters  \cite{wiseman2009quantum}.  Recent interest in this field has moved beyond simple multiplicative parameters of the Hamiltonian and begun to examine general parameters \cite{pang2014quantum}, as well as the role of physical dynamics in the estimation process \cite{liu2015quantum,PhysRevA.92.012312}, which may require coherent control to optimize the acquired information \cite{PhysRevLett.115.110401,pang_quantum_2016}.

The purpose of this Letter is to apply the methods of quantum metrology to the LZ transition, and quantify the estimation precision of parameters in the LZ transition  available by the various techniques aforementioned. We shall focus on the quantum Fisher information for the parameters of interest, as it determines the lower bound of the variance of the parameter estimates over all possible estimation strategies and all possible quantum measurements on the systems, giving the ultimate limits of precision allowed by quantum mechanics in the asymptotic data limit. We find that because of the time-dependent nature of the problem, with a proper control Hamiltonian applied, the time-scaling of the quantum Fisher information can be significantly improved, which demonstrates a fundamental metrological advantage of coherent quantum control on the level-crossing physics of the LZ transition.

The LZ Hamiltonian is given by
\begin{equation}
H(t)=\dfrac{vt}{2}\sigma_{z}+\dfrac{\Delta}{2}\sigma_{x}, \label{eq:H}
\end{equation}
where $v$ is the speed of the sweep, $\Delta$ is the level splitting at the transition time $t=0$. Denote the solution to the Schr\"odinger equation, $i \partial_t |\psi\ra = H(t) |\psi\ra$ as $\ket{\psi(t)}=C_{0}(t)\ket{0}+C_{1}(t)\ket{1}$, which gives two coupled differential equations for $C_{0,1}(t)$. Eliminating $C_1$ transforms the equation for $C_0$ into the Weber equation, solved by parabolic cylinder functions \cite{whittaker_course_1996,wang_special_1989}.  
We start for simplicity in the ground state $|1\ra$ at an initial time $t=-T_0$ far away from the avoided level crossing time $t=0$, i.e., 
$
T_0\gg \tau\equiv \max \{\frac{\Delta}{2v},\frac{1}{\sqrt{v}}\}
$. 
Sweeping through the Landau-Zener transition to a time $t=T \gg \tau$, 
which is also far away from the transition region (see Fig.~1 inset), justifies the asymptotic expansions of the parabolic cylinder functions to give
\begin{equation}
C_{0}(T)=\dfrac{\sqrt{2\pi i \gamma}}{\Gamma(1+\nu)}e^{-\pi \gamma/2 -2 i\phi}, \
C_{1}(T) = e^{-\pi \gamma}, \label{eq:Cs}
\end{equation}
where we define $\phi \equiv (v T^{2}+\pi)/4+\gamma /2 \ln(v T^2)$ and $\gamma = \Delta^{2}/(4v)$ \cite{SM}.
The absolute square of $C_{0,1}(T)$ recovers the celebrated (time-independent) LZ probabilities \cite{zener_non-adiabatic_1932} to find the system in the (new) excited or ground states,
\begin{equation}
P_{1}= 1 - P_0 = \big|C_{1}(T)\big|^{2}=e^{-2\pi \gamma}.\label{eq:C2TSQ}
\end{equation}

{\it Estimation using the LZ probabilities.}---The simplest estimation scheme is to make a single pass starting from the ground state $|1\ra$, and measure the system to be in the new excited or ground state, with the probabilities given in (\ref{eq:C2TSQ}). Since the probabilities depend very sensitively on the parameters $v$ or $\Delta$ in the Hamiltonian (\ref{eq:H}), they may be estimated with those probabilities according to classical estimation theory, with an unbiased estimator whose variance is bounded by the inverse of the Classical Fisher information (CFI) of the parameter $g$ (the Cram\'er-Rao bound \cite{cramer1946mathematical}), given by
$F_{g}=\sum_{\xi}\frac{1}{p(\xi|g)}[\frac{\partial p(\xi |g)}{\partial g}]^{2} 
$, where $\xi=0,1$. The corresponding CFIs at time $T$ for $g=v$, and for $g=\Delta$ are \cite{SM},
\begin{equation}
F_{\Delta}(T)=\dfrac{16\pi^{2}\gamma^{2}}{(e^{2\pi \gamma}-1)\Delta^{2}}, \quad
F_{v}(T)=\dfrac{4\pi^{2}\gamma^{2}}{(e^{2\pi \gamma}-1)v^{2}}. \label{eq:Fv}
\end{equation}

Repeating the experiment $N$ times from the same initial state will boost the information by a factor of $N$.
The Fisher information about either parameter limits to zero for either a diabatic transition $\gamma \ll 1$, or an adiabatic transition $\gamma \gg 1$.  This is simply because in those extreme limits, the LZ probabilities become either 0 or 1, with little variation.  Therefore, the strategy is most sensitive in the intermediate range.  For $\gamma$ of order 1, the uncertainty of both parameters is of order of the parameter, which for tiny tunnel couplings can give rise to precise estimates \cite{wernsdorfer2000landau}. 

\begin{figure}
\begin{centering}
\includegraphics[width=\columnwidth]{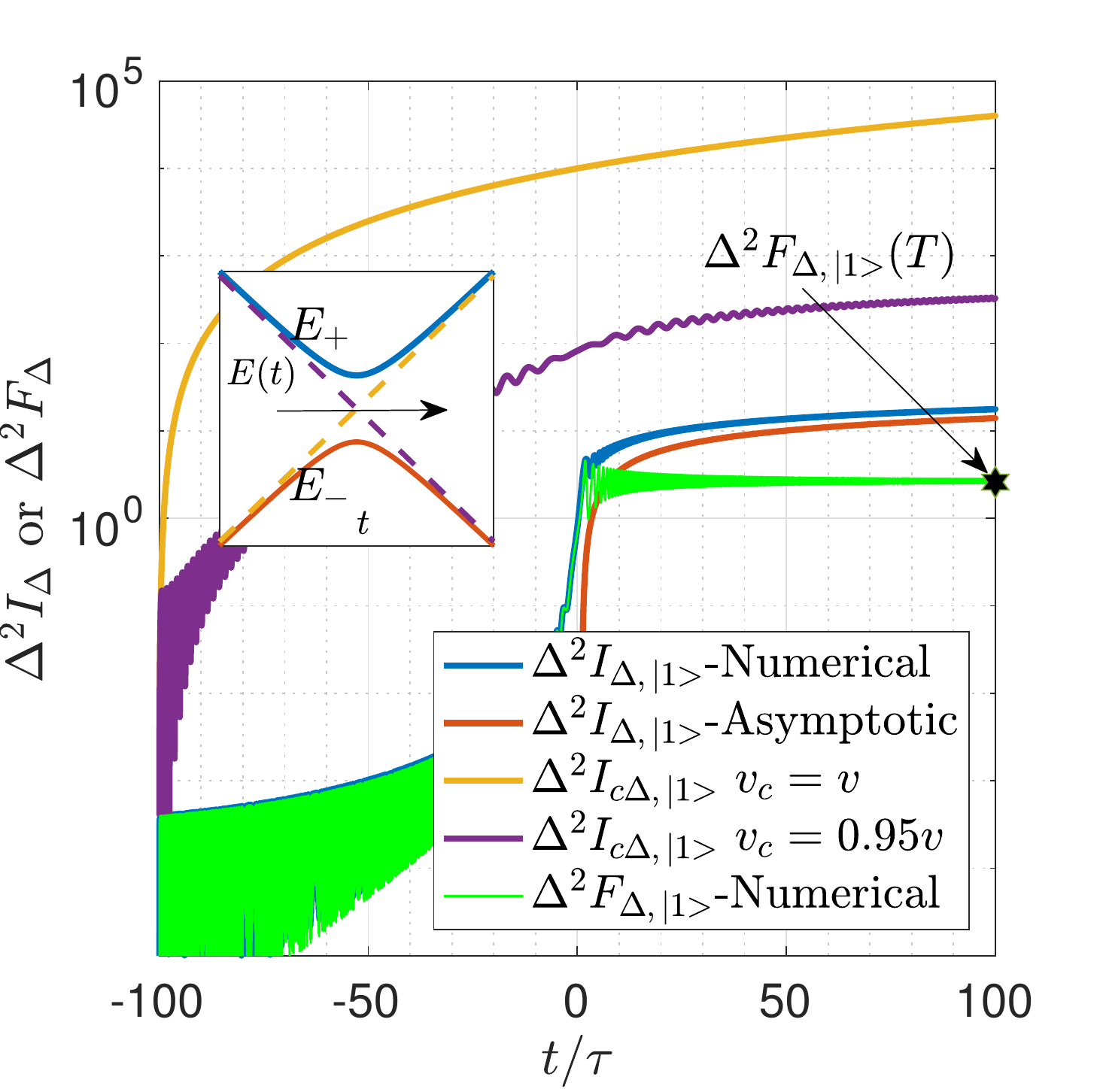}
\par\end{centering}
\caption{QFIs/CFIs for estimating $\Delta$ versus
time plotted in logarithmic graph(base $10$). $I_\Delta$ or $F_\Delta$ denote the QFI or CFI without control while  $I_{c\Delta}$ denotes the QFI with control. The value of parameters in the LZ Hamiltonian are $v=1$, $\Delta=1$, and $\tau=1$. The system starts to evolve at $-T_0=-100\tau$. For cases with control Hamiltonians, we choose the initial state to be $\frac{1}{\sqrt{2}}\left[\ket{+x}-\ket{-x}\right]=\ket{1}$. The black marker is the result calculated from Eq.~(\ref{eq:Fv}). The inset represents a single LZ transition.} \label{fig:FIDel}
\end{figure}

\begin{figure}[t]
\begin{centering}
\includegraphics[width=\columnwidth]{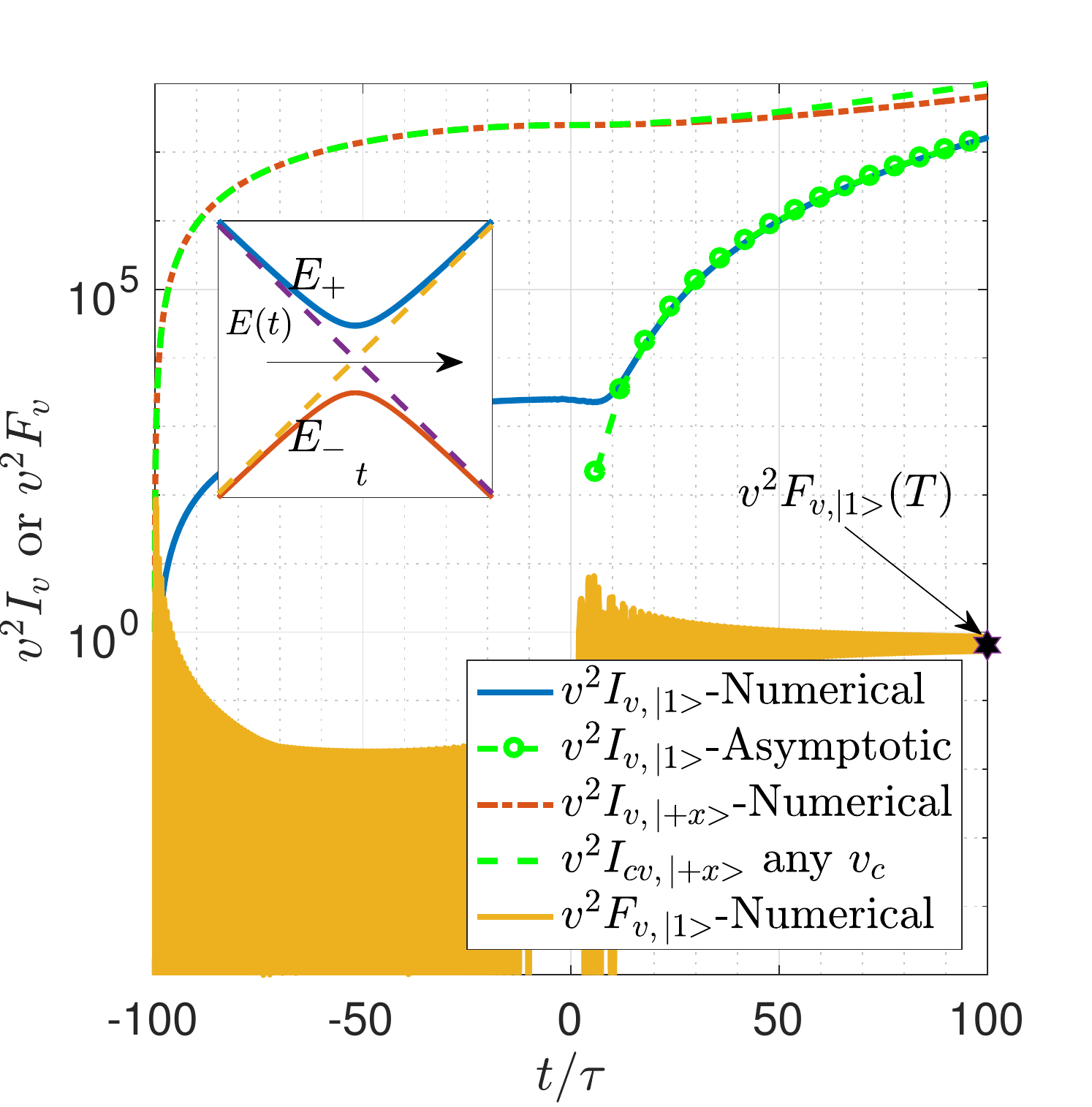}
\par\end{centering}
\caption{QFIs/CFIs for estimating $v$ versus time
in a semi logarithmic plot (base $10$). $I_v$ or $F_v$ denote the QFI or CFI without control while  $I_{cv}$ denotes the QFI with control. The parameter configurations are the same as FIG.~\ref{fig:FIDel}. The green dashed line is the case with optimal controls, where $H_{c}=-\frac{\Delta}{2}\sigma_{x}$ and $v_c$, the control parameter in the OLCH, can be arbitrarily chosen. The black marker is the result calculated from Eq.~(\ref{eq:Fv}). The inset represents a single LZ transition.} \label{fig:FIv}
\end{figure}

{\it Estimation using any final quantum measurement.}--- We can generalize the above situation by rather than making a final measurement at time $T$ in the $|0\ra$, $|1\ra$ basis, to measure in another basis (or equivalently, stopping the LZ sweep and applying a single qubit unitary).  The 
maximum Classical Fisher Information over all possible generalized quantum measurements on a state $\ket{\psi_{g}}$ is defined as the Quantum Fisher Information (QFI) \cite{braunstein_statistical_1994,braunstein_generalized_1996,paris_quantum_2009},
$I_{g}=4(\braket{\partial_{g}\psi_{g}\big|\partial_{g}\psi_{g}}-|\braket{\psi_{g}\big|\partial_{g}\psi_{g}}|^{2})$.
It has been shown \cite{paris_quantum_2009} that the optimal measurements associated with the QFI are projective measurements formed by the eigenvectors of $L_{g}$, defined as $L_{g}=2\partial_{g}\rho_{g}=2\left(\ket{\partial_{g}\psi_{g}}\bra{\psi_{g}}+\ket{\psi_{g}}\bra{\partial_{g}\psi_{g}}\right)$.
For the two-level Landau-Zener model, the matrix elements of $L_{g}$ operator in the $ |0\ra,|1\ra$ basis becomes $[L_{g}(t)]_{ij}=2\partial_{g}\left[C_{i}(t)C_{j}^{*}(t)\right](i,j=0,1)$.
The QFI gives better precision since it is able to take advantage of the phase that is rapidly accumulating during the LZ sweep, $\varphi(T) \propto v T^2$, as predicted by Stueckelberg \cite{stueckelberg1932theory}.  Still starting from the ground state at $t=-T_0$, as one can observe from Eq.~(\ref{eq:Cs},\ref{eq:C2TSQ}), the state at time $T$ can be rewritten as: $|\psi(T)\ra=\sqrt{P_0}|0\ra+\sqrt{P_1} e^{i \varphi(T)}|1\ra$,
where the relative phase is
$\varphi(T)=\frac{v T^2}{2}+\gamma \ln(v T^2)+\text{arg} \Gamma(1-\gamma)+\frac{\pi}{4}$. 
When $T$ is sufficiently large, one can calculate the QFIs by keeping only the contributions due to highest order of $T$ in the relative phase and neglecting the contributions from the transition probabilities (for a rigorous treatment, see \cite{SM}) as follows
\begin{equation}
I_{\Delta}(T)\sim\frac{\Delta^{2}}{v^{2}}P_{0} P_{1} \left[\ln\left(vT^{2}\right)\right]^{2}, \quad
I_{v}(T)\sim P_{0} P_{1} T^{4}. \label{eq:Ieps0T}
\end{equation}
The above prefactors $P_0 P_1$ attains its maximum $1/4$ when the transition probabilities are equal.  Both of the QFIs exceed the CFIs, with $I_v$ scaling as $T^4$ because the acquired phase difference scales as $T^2$ (note this time scaling originates in the explicit linear time growth of the Hamiltonian (\ref{eq:H}), which is different from the case in \cite{pang_quantum_2016}). In general, we may further boost the QFIs by starting the system in a coherent superposition of $|0\ra$ and $|1\ra$, however, while this effects the prefactor of the QFIs, it does not change the time scaling.  The vectors forming the optimal projectors, either for estimation of $\Delta$ or $v$, in the $\sigma_z$ basis, can also be immediately obtained as an equal superposition of $\ket{0}$ and $\ket{1}$ with relative amplitudes
$\pm i e^{i \varphi(T)}$ \cite{SM}.
Plots of the CFIs and QFIs are shown in Figs.~(\ref{fig:FIDel},\ref{fig:FIv}) for $g=\Delta,\,v$ respectively. 
Although the previous discussion assumes a positive sweeping velocity starting from the ground state, the discrete symmetries of the LZ Hamiltonian \eqref{eq:H}, relate this solution to the negative case velocity and to starting in the excited state; all these cases have the same CFIs or QFIs and the corresponding optimal measurements \cite{SM}.

{\it Adding coherent control to boost precision.}---It has been pointed out \cite{pang_quantum_2016} that for a general
time dependent Hamiltonian the QFI at time $t$ is bounded by $I_{cg}(t)\leq
[\intop_{t_{0}}^{t}\left(\mu_{\max}(t^{\prime})-\mu_{\min}(t^{\prime})\right)dt^{\prime}]^{2}$, 
where the subscript $c$ denotes the QFI with coherent controls; $t_{0}$ is the initial time of the evolution of the system; and $\mu_{\max}(t)$ and $\mu_{\min}(t)$ are the maximum instantaneous eigenvalues of
$\partial_{g}H_g(t)$. The equality can be saturated if the initial
state is prepared in the superposition of the maximum and minimum
eigenstates of $\partial_{g}H_g$, where the maximum (minimum) eigenstate denotes the eigenstate corresponding to the maximum (minimum) eigenvalue of $\partial_{g}H_g$, and an Optimal Control Hamiltonian (OCH) is applied of the form
$H_{c}\left(t\right)=\sum_{k}f_{k}\left(t\right)\ket{\psi_{k}(t)}\bra{\psi_{k}(t)}-H_{g}(t) +i\sum_{k}\ket{\partial_{t}\psi_{k}\left(t\right)}\bra{\psi_{k}\left(t\right)}$,
where $\ket{\psi_{k}(t)}$ is the $k$th eigenstate of $\partial_{g}H_{g}$; $f_{k}(t)$ can be  taken arbitrary in principle, but is usually chosen to take the form which simplifies the OCH $H_{c}\left(t\right)$ significantly. 

If the system is prepared in an eigenstate of
$\partial_{g}H_{g}$ initially, the functionality of the OCH is to steer quantum state, such that the system remains
in the eigenstate of $\partial_{g}H_{g}$ under time evolution with $H_g(t)+H_c(t)$. However,  if
a level crossing occurs at some time point  between maximum (minimum)
states with other eigenstates of $\partial_{g}H_{g}$, where we denote
the old maximum(minimum) state before level crossing as $\ket{\psi_{n}}$
and the new maximum(minimum) state after level crossing as $\ket{\psi_{m}}$,
in order to achieve the maximum QFI, an additional Optimal Level Crossing Hamiltonian (OLCH) $H_{LC}(t)$ is required to rotate from $\ket{\psi_{n}}$
to $\ket{\psi_{m}}$, which will be important here because in the case of $g=v$, a level crossing occurs in $\partial_v H_v$.  The general expression of  $H_{LC}$ as well as its applications to current single LZ transition and the periodic LZ transitions discussed later are included in the Supplemental Material \cite{SM}. 

Applying this theory of time-dependent quantum metrology to estimate
$\Delta$, we find $\partial_{\Delta}H=\sigma_{x}/2$, with eigenvalues $\pm 1/2$. Applying the above results for the QFI with respect to $\Delta$, we find the upper bound 
\begin{equation}
I_{c\Delta,\,\ket{\Psi}_{c\Delta}}(t)= \Big( \int_{-T}^t dt^{\prime}\Big)^2 = (t+T)^{2}\label{eq:maxIDeltCtrl},
\end{equation}
for any time $t$, and giving a maximum of $4 T^2$ at $t=T$, provided the initial state is prepared in $\ket{\Psi}_{c\Delta}=(1/\sqrt{2})[\ket{+x}+e^{i\beta}\ket{-x}]$, 
where $\beta$ is an arbitrary initial relative phase.
The corresponding OCH $H_{c}=-(vt/2)\sigma_{z}$
cancels the first term in Eq.~(\ref{eq:H}), effectively turning off the LZ sweep in $\sigma_z$. In constructing
the optimal control Hamiltonian, we have taken $f_{\ket{\pm x}}=\pm\Delta/2$. No OLCH is required since $\partial_{\Delta}H$  and its eigenstates are time
independent and no level crossing occurs. Note that if $v$ is unknown, we should replace $v$ in the OCH with an estimate $v_c$ that can be updated based on further measurement data. Fig.~\ref{fig:FIDel} shows the comparison of the optimal case with the non-control and non-optimal cases.

The estimation of $v$ with control is more complicated than $\Delta$ since the maximum and minimum eigenvalues of $\partial_{v}H=t\sigma_{z}/2$
have a crossing at $t=0$. The QFIs for all time can be written in a uniform expression 
\begin{equation}
I_{cv,\,\ket{\Psi}_{cv}}(t)=\Big(\int_{-T}^{t} \big|t^{\prime} \big|  d t^{\prime} \Big)^{2}=\dfrac{[t^{2}+\text{sgn}(t) T^{2}]^{2}}{4},
\end{equation}
where the value of $\text{sgn}(t)$ is $-1$ for $t\le 0$ and $1$ for $t>0$. We  prepare the initial state in $\ket{\Psi}_{cv}=\left[\ket{0}+e^{i\beta}\ket{1}\right]/\sqrt{2}$ with an arbitrary chosen relative phase $\beta$.
Taking $f_{\ket{0}}=vt/2$ and $f_{\ket{1}}=-vt/2$, the OCH becomes $H_{c}=-(\Delta/2)\sigma_{x}$, which cancels the tunneling term. Since the maximum
and minimum eigenstates of $\partial_{v} H$ have a level crossing at $t=0$, an OLCH $H_{LC}$ is required to avoid the level crossing of $\partial_v H$
at $t=0$. This can be done simply by swapping the eigenstates of $\sigma_z$ with a $\pi$-pulse at time $t=0$.  An explicit construction is given in the Supplemental Material \cite{SM}, which is valid even if the estimate of $v$ is imperfect ($v_c\neq v$). The comparison of the optimal case with other cases are plotted in Fig.~\ref{fig:FIv}. 

{\it Optimal measurements.}---In order to saturate the bounds with optimal controls, it is necessary to construct the optimal measurements. For estimating $\Delta$, if the OCH is applied and the system is initially prepared in $\ket{\Psi}_{c\Delta}$,
the system will evolve under the Hamiltonian $H+H_{c}=\frac{\Delta}{2}\sigma_{x}$.
The vectors forming the corresponding optimal projectors, expressed in the $\sigma_x$ basis, are equal superposition of $\ket{+x}$ and $\ket{-x}$ with relative phases $\pm i e^{i \Delta(t+T)+\beta}$  (see \cite{SM} for details). 
For estimating $v$ with optimal controls applied, similar arguments give rise to vectors forming the  measuring projectors, expressed in the $\sigma_z$ basis, are equal superposition of $\ket{0}$ and $\ket{1}$ with relative amplitudes $\pm i e^{i\frac{v_c}{2}(t^{2}-T^{2})+\beta}$ for $t<0^{-}$ and $\pm i e^{i \frac{v_c}{2}(t^{2}+T^{2})-\beta-v_{c}T^{2}}$ for $t>0^{+}$, where $v_c$ is the control parameter appearing in the OLCH \cite{SM}.

{\it Optimal estimation with controlled LZ interferometry.}---Rather than take a single pass though the avoided level crossing, the concept of LZ interferometry is to make many passes, acquiring a phase shift given by a multiple of the phase shift acquired by a single cycle \cite{shytov2003landau,oliver2005mach,shevchenko2010landau}.  This leads to interference fringes in the occupation probability, known as ``Stueckelberg oscillations'' \cite{PhysRevLett.68.3515,PhysRevLett.69.1919,nakamura2002concepts}.  In contrast to past work, we will see that simply letting the phase accumulate does not give the optimal precision.  Rather a series of control operations should be applied to optimize the information extraction and change the scaling law of the Fisher information with duration.  This situation allows us to extend the time $T$ of the experiment and gives an explicitly bounded Hamiltonian, in contrast to a single sweep, where the LZ Hamiltonian approximation (\ref{eq:H}) would otherwise break down at long time.
\begin{figure}[tbh!]
\begin{centering}
\includegraphics[width=\columnwidth]{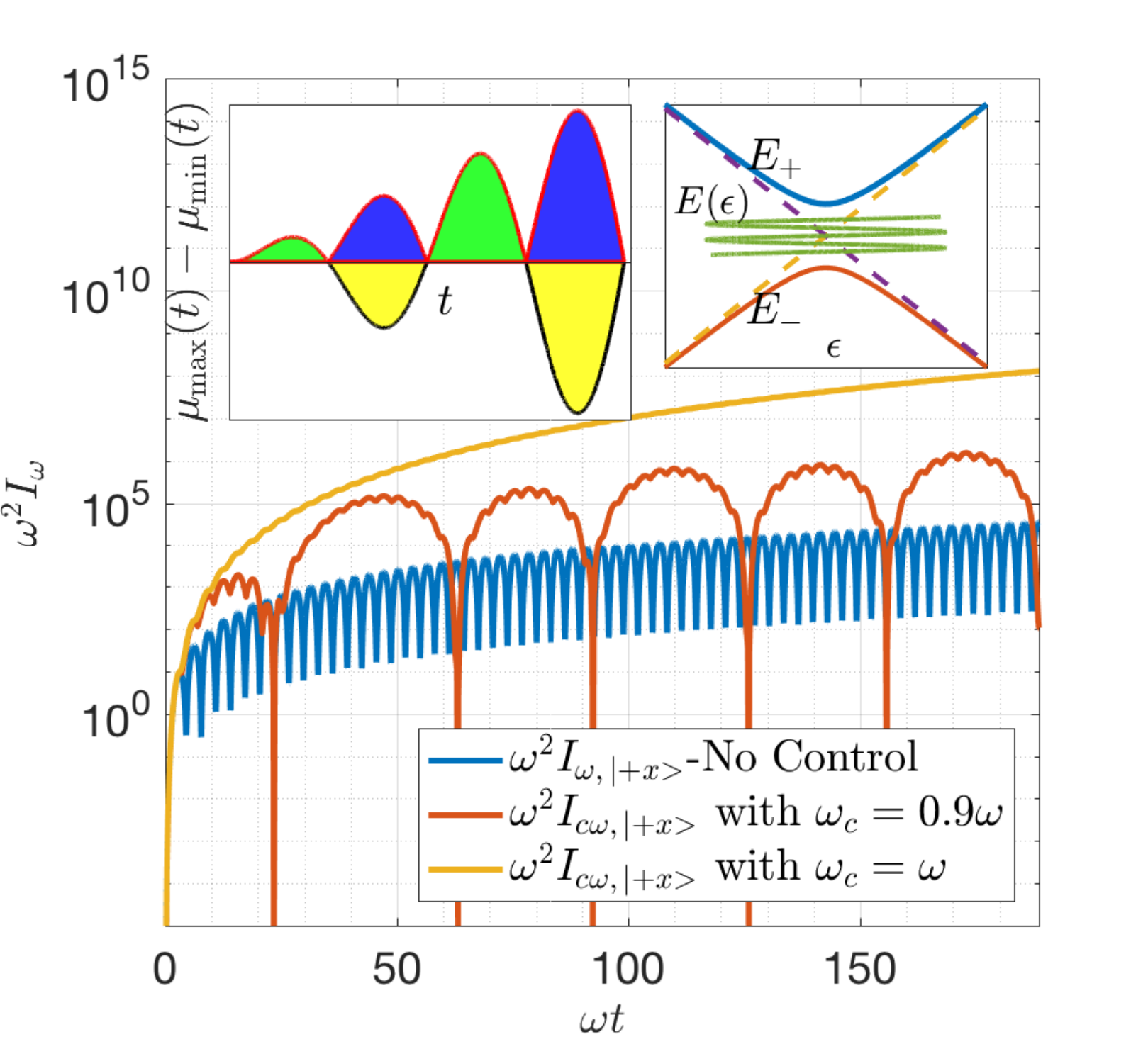}
\par\end{centering}
\caption{The main figure is the QFIs for estimating $\omega$ versus
time in semi-logarithmic graph (base $10$). The subscript $c$ in the notation of QFI implies it corresponds to the case with controls. The system starts
to evolve at $t=0$ and the value of parameters are $\epsilon_{0}=0$,
$A=1$, $\omega=1$, $\Delta=0.1$, $N=60$, $T=\frac{N\pi}{\omega}$=$60\pi$. The two cases with controls have the same OCH $H_{c}=-\frac{\epsilon_0}{2}\sigma_{z}-\frac{\Delta}{2}\sigma_{x}$, whereas the additional control Hamiltonian is not optimal (yellow) and optimal (purple). The left inset shows the merit of the OLCHs: consider two cases both initially prepared in $(\ket{0}+\ket{1})/\sqrt{2}=\ket{+x}$ and both with OCHs applied; the one without OLCHs has $\sqrt{I_{c\omega}}=\big|G-Y\big|$; the other with OLCHs has $\sqrt{I_{c\omega}}=G+B$, where $G$, $Y$, $B$ represent the magnitudes of the green, yellow and blue areas.  The right inset represents oscillatory avoided level crossings.
\label{fig:Iomg}}
\end{figure}

The Hamiltonian (\ref{eq:H}) is modified by replacing the coefficient of $\sigma_z$ by an oscillating function of amplitude $A$ and frequency $\omega$, $v t \rightarrow \epsilon_0+A \cos \omega t$, describing periodic LZ sweeps.  We restart our clock from $t=0$, beginning away from the transition region.   We can now estimate four Hamiltonian parameters, but we focus on the frequency $\omega$ as the most interesting.  A direct solution of the problem and calculating the associated QFIs is rather involved. However, when $\epsilon_0=0$, $A \ll \Delta, \omega$ and $|\omega-\Delta|\ll|\omega+\Delta|$  (weak coupling and near resonance), by making two consecutive transformations $\ket{\psi(t)}_{E}=U_E \ket{\psi(t)}$, where $U_E \equiv e^{i\Delta/2\sigma_{z}t}e^{i \sigma_{y} \pi/4}$ and $\psi(t)$ is the state corresponding to the Hamiltonian $\frac{1}{2}A\cos\left(\omega t\right)\sigma_{z}+\frac{1}{2}\Delta\sigma_{x}$ in the original lab frame, then applying the rotating wave approximation \cite{SM,grifoni_driven_1998}, the transformed Hamiltonian is then
\be
\mathcal{H}_{E}=-\dfrac{A}{4}\left\{ \cos\left[(\omega-\Delta)t\right]\sigma_{x}-\sin\left[(\omega-\Delta)t\right]\sigma_{y}\right\}.
\ee
Since $U_E$ does not depend on the estimation parameter $\omega$, the QFI in the original lab frame is identical to that in the transformed frame as one can verify straightforward by the definition of the QFI. The maximum  QFI  at $t=T$ over all possible initial pure states of estimating $\omega$ with $\mathcal{H}_{E}$ is given in \cite{pang_quantum_2016}, and scales as $T^2$; $\max I_{\omega}(T)=A^{2}T^{2}/(A^{2}+4\left(\Delta-\omega\right)^{2})$, plus oscillatory terms of sub-leading order. The analytic treatment of the cases of strong coupling and off resonance is rather involved, so a numerical simulation for these cases is presented in Fig.~\ref{fig:Iomg}. However, by adding optimal controls, we can find the QFI of estimating $\omega$ for general cases, regardless of  the driven intensity and frequency.  We find the parametric derivative of the Hamiltonian is $\partial_\omega H = - A t \sin (\omega t) \sigma_z/2$, which has eigenvectors $|0\ra, |1\ra$, and eigenvalues $\mu_{\mp} = \mp A t \sin (\omega t) /2$.  An interesting feature arises in that there is a crossing of the eigenvalues of $\partial_\omega H$ at the {\it ends} of the LZ sweeps, not at the crossing of the energy eigenvalues.  Additional OLCHs must be applied at each of these time points to swap the amplitudes of $|0\ra$ and $|1\ra$ in order to saturate the quantum Fisher information bound. Since the oscillation frequency is not precisely known, generally the controls are applied with an estimated value $\omega_c$, which is then iteratively updated in successive trials \cite{pang_quantum_2016}.
The left inset of Fig.~\ref{fig:Iomg} schematically shows the functionality of the OLCHs: when OCH is applied, the square root of the QFI is the integrated difference of the maximum and minimum eigenvalues of $\partial_\omega H$ over time, which in the absence of OLCHs is the magnitude of the difference of the green and yellow areas, whereas in presence of the OLCHs is the sum of green and blue areas. With all optimal controls applied, the QFI is
\begin{equation}
I_{c\omega,\ket{\Psi}_{c\omega}}=\Big(\dfrac{A  \pi N^{2}}{\omega^{2}}\Big)^2,
\end{equation}
where we have considered $T=N\pi/\omega, (N\in\mathbb{N})$ for simplicity (the general solution is given in \cite{SM}) and the system is initially prepared in $|\Psi\ra_{c\omega}=\frac{1}{\sqrt{2}}\left(\ket{0}+e^{i\beta}\ket{1}\right)$ with an arbitrary initial relative phase  $\beta$.  We see that the QFI scales as $T^4$, giving a scaling law improvement in the estimation of $\omega$.  The required OCH is $H_c = -\epsilon_0 \sigma_{z}/2- \Delta \sigma_x/2$ applied in addition to the OLCHs.  A comparison of the optimal case with both the non-control and non-optimal case is plotted in the main figure of Fig.~\ref{fig:Iomg}. The  vectors forming the optimal projective measurements for estimation of $\omega$, written in the $\sigma_z$ basis, are equal superposition of $\ket{0}$ and $\ket{1}$ with relative amplitude $\pm i^{N+1}(-1)^{Nl+N} e^{i(-1)^{N}\beta}$, where $l$ is an integer appearing in the OLCHs \cite{SM}.

The essential difference between the $T^4$ scaling in this case and the one in the single sweep case is worth noting. For LZ interferometry, the Hamiltonian is bounded in time, so the quantum Fisher information cannot be simply increased by the time growth of the Hamiltonian. The quantum control-enhanced time scaling of Fisher information still comes from the time-dependence of the Hamiltonian, since the acquired phase accelerates in time, which leads to the $T^4$ scaling of Fisher information, similar to \cite{pang_quantum_2016}.
 
{\it Conclusions.}--- The physics of the Landau-Zener transition depends very sensitively on the parameters of the underlying Hamiltonian.  We have quantified the ultimate precision allowed by quantum mechanics based on the preparation, evolve for a given time, and measure paradigm, using the quantum Fisher information metric.  By building up from using the LZ transition probabilities, to the acquired phase, to multiple transitions, we have shown that increasing precision may be obtained.  By further applying coherent quantum control together with adaptive feedback, the ultimate limits of time-dependent quantum metrology may be achieved, and we demonstrated the $T^4$ scaling of the quantum Fisher information for the oscillation frequency.  We have given an explicit measurement prescription to unlock the additional quantum advantages in the measurement time resource, and our numerical simulations have confirmed the analytic results.

{\it Acknowledgments.}---
This work was supported by 
by US Army Research Office Grants No. W911NF-15-1-0496, No. W911NF-13-1-0402, and by National Science Foundation grant DMR-1506081. 

\bibliographystyle{apsrev4-1}

\newpage \setcounter{equation}{0} \setcounter{subsection}{0} \renewcommand{\theequation}{S\arabic{equation}} \onecolumngrid \setcounter{enumiv}{0}

\section*{Supplemental Material}

\section{Background of the Quantum Fisher Information\label{sec:BackgroundQFI}}

In this section we present some background information in the Quantum
Fisher Information(QFI), on which the main text is based.

It has been shown that the maximum Classical Fisher Information(CFI)
over all possible generalized quantum measurements (known as positive
operator valued measurements) on a pure state $\ket{\psi_{g}}$ is
defined as the Quantum Fisher Information(QFI) \cite{braunstein_statistical_1994,braunstein_generalized_1996,paris_quantum_2009}
\begin{equation}
I_{g}=4\left(\braket{\partial_{g}\psi_{g}\big|\partial_{g}\psi_{g}}-\Big|\braket{\psi_{g}\big|\partial_{g}\psi_{g}}\Big|^{2}\right),\label{eq:Ig-def}
\end{equation}
Note that Eq. \eqref{eq:Ig-def} is always positive since one can
prove by Cauchy-Schwarz inequality $\braket{u\big|u}\braket{v\big|v}\geq\big|\braket{u\big|v}\big|^{2}$
with $\ket{u}=\ket{\partial_{g}\psi_{g}}$ and $\ket{v}=\ket{\psi_{g}}$.
It has been shown that \cite{paris_quantum_2009}, for a general state
$\rho_{g}$, where $\rho_{g}$ is the density operator, the optimal
observable to give rise to the QFI is
\begin{equation}
O_{g}=g\mathbb{I}+\dfrac{L_{g}}{I_{g}},
\end{equation}
where $L_{g}$ is the Symmetric Logarithmic Derivative(SLD) defined
as
\begin{equation}
\dfrac{L_{g}\rho_{g}+\rho_{g}L_{g}}{2}=\partial_{g}\rho_{g}.
\end{equation}
The optimal measurements associated with Eq. \eqref{eq:Ig-def} can
be found to be the projective measurements formed by the eigenvectors
of $O_{g}$, which are also eigenvectors of $L_{g}$. Particularly,
if the state is pure $\rho_{g}=\ket{\psi_{g}}\bra{\psi_{g}}$ and
$\rho_{g}^{2}=\rho_{g}$, we have $\partial_{g}\rho_{g}=\partial_{g}\rho_{g}\rho_{g}+\rho_{g}\partial_{g}\rho_{g}$.
Thus for this case 
\begin{equation}
L_{g}=2\partial_{g}\rho_{g}=2\left(\ket{\partial_{g}\psi_{g}}\bra{\psi_{g}}+\ket{\psi_{g}}\bra{\partial_{g}\psi_{g}}\right).
\end{equation}
In particular, if $\ket{\psi_{g}}=\sqrt{P_{0}}\ket{0}+\sqrt{P_{1}}e^{i\phi_{g}}\ket{1}$
where $P_{0}$ and $P_{1}$ are real and independent of $g$, the
QFI is
\begin{equation}
I_{g}=4P_{0}P_{1}\left(\partial_{g}\phi_{g}\right)^{2}\label{eq:Ig-special}
\end{equation}
and the operator$L_{g}$ in the $\ket{0}$, $\ket{1}$ basis becomes
\begin{equation}
L_{g}(t)=2\partial_{g}\phi_{g}P_{0}P_{1}\left[\begin{array}{cc}
0 & -ie^{-i\phi_{g}}\\
ie^{i\phi_{g}} & 0
\end{array}\right].\label{eq:Lg-1}
\end{equation}
The corresponding eigenvectors in the $\ket{0}$, $\ket{1}$ basis
are 
\begin{equation}
\ket{\pm}_{g}=\frac{1}{\sqrt{2}}\left(\ket{0}\pm ie^{i\phi_{g}}\ket{1}\right),\label{eq:pm}
\end{equation}
 where we have used the fact that removing the prefactor $2\partial_{g}\phi_{g}$$P_{0}P_{1}$
does not change the eigenvectors of $L_{g}$. Eq. \eqref{eq:pm} means
the the directions of two optimal projective measurements only depend
on the relative phase if the transition probabilities are independent
of $g$.

In the main text, we find for a system evolving under the Landau-Zener
(LZ) Hamiltonian \eqref{eq:H-1} from the ground state, the state
at the end of the LZ transition when $T$ is sufficiently large is
\begin{equation}
\ket{\psi(T)}=\sqrt{P_{0}}\ket{0}+\sqrt{P_{1}}e^{i\varphi(T)}\ket{1},\label{eq:psiT}
\end{equation}
where $\ensuremath{\varphi(T)=\frac{vT^{2}}{2}+\gamma\ln(vT^{2})+\text{arg}\Gamma(1-\gamma)+\frac{\pi}{4}}.$
According to Eq. \eqref{eq:pm}, the corresponding optimal measurements
are formed by 
\begin{equation}
\ket{\pm}_{g}=\frac{1}{\sqrt{2}}\left\{ \ket{0}\pm ie^{i\varphi(T)}\ket{1}\right\} ,
\end{equation}
where $g=\Delta$ or $g=v$.

\section{Symmetries of the Landau-Zener (LZ) Hamiltonian\label{sec:SymmetriesLZ}}

In this section, we prove that the following four cases have the same
CFI or QFI: $(1)$ positive transition velocity starting from ground
state; $(2)$ negative transition velocity starting from excite state;
$(3)$ positive transition velocity starting from excited state; $(4)$
negative transition velocity starting from ground state. 

The LZ Hamiltonian is 
\begin{equation}
H_{v}=\dfrac{vt}{2}\sigma_{z}+\dfrac{\Delta}{2}\sigma_{x}.\label{eq:H-1}
\end{equation}
We denote the QFI for the four cases as $I_{v,\,\ket{1}}$, $I_{-v,\,\ket{0}}$,
$I_{v,\,\ket{0}}$, $I_{-v,\,\ket{1}}$. We first prove $I_{v,\,\ket{1}}=I_{-v,\,\ket{0}}$
and $I_{v,\,\ket{0}}=I_{-v,\,\ket{1}}$ by applying $\sigma_{x}$
transformation to the state. 

\textit{Proof: }Assume the system is in the case $(1)$ condition,
i.e, initially in the ground state $\ket{1}$ and the transition velocity
in Eq. \eqref{eq:H-1} is positive. The solution the Schrödinger equation
corresponding to this case is denoted as
\begin{equation}
\ket{\psi(t)}=C_{0}(t)\ket{0}+C_{1}(t)\ket{1}.\label{eq:psi-t}
\end{equation}
Applying the transformation $\ket{\hat{{\psi}}(t)}=\sigma_{x}\ket{\psi(t)}$,
we have
\begin{equation}
i\partial_{t}\ket{\hat{\psi}(t)}=\sigma_{x}H_{v}\sigma_{x}\ket{\hat{{\psi}}(t)}=H_{-v}\ket{\hat{\psi}(t)},
\end{equation}
where $H_{-v}$ is the LZ Hamiltonian \eqref{eq:H-1} with $v$ replaced
by $-v$. The initial state then transforms from $\ket{1}$ to $\sigma_{x}\ket{1}=\ket{0}$.
Thus, we know if the system is in condition of case (2), its state
would be 
\begin{equation}
\ket{\hat{\psi}(t)}=\sigma_{x}\ket{\psi(t)}=C_{1}(t)\ket{0}+C_{0}(t)\ket{1}.\label{eq:psi-hat}
\end{equation}
Plugging Eqs. (\ref{eq:psi-t}, \ref{eq:psi-hat}) into the expression
for QFI, one concludes that $I_{v,\,\ket{1}}=I_{-v,\,\ket{0}}$. Following
from the same arguments, one can also prove $I_{v,\,\ket{0}}=I_{-v,\,\ket{1}}$. 

Next, we prove that $I_{v,\,\ket{1}}=I_{v,\,\ket{0}}$ and $I_{-v,\,\ket{1}}=I_{-v,\,\ket{0}}$
by applying the $-i\sigma_{y}K$ operation to the state, where $K$
is the complex conjugate operator defined as $K\ket{\psi}=C_{0}^{*}\ket{0}+C_{1}^{*}\ket{1}$.
In fact, this is the time reversal operation for spin $\frac{1}{2}$
system. Applying the transformation $\ket{\widetilde{\psi}(t)}=-i\sigma_{y}K\ket{\psi(t)}$,
the Schödinger equation becomes
\begin{equation}
i\partial_{t}\ket{\widetilde{\psi}(t)}=\left(-i\sigma_{y}K\right)H_{v}\left(-i\sigma_{y}K\right)^{-1}\ket{\widetilde{\psi}(t)}=\left(-i\sigma_{y}\right)KH_{v}K\left(i\sigma_{y}\right)\ket{\widetilde{\psi}(t)},
\end{equation}
where we have used $K^{-1}=K$ and $\left(-i\sigma_{y}\right)^{-1}=i\sigma_{y}$.
In the $\ket{0}$, $\ket{1}$ basis, we have 
\begin{equation}
\text{i}C_{i}^{*}=-\sum_{j=0,1}[H]_{ij}^{*}C_{j}^{*},\label{eq:ccSchEq}
\end{equation}
which following from taking the complex conjugate on both sides of
the Schödinger equation. Therefore, by comparing Eq. \eqref{eq:ccSchEq}
with $iK\ket{\psi}=\left(KHK\right)K\ket{\psi}$, we conclude that
$KHK=-H^{*}$, where the complex conjugate of $H$ means taking the
complex conjugate of each matrix element of $H$ in the $\ket{0}$,
$\ket{1}$ basis. Since the matrix elements of the LZ Hamiltonian
\eqref{eq:H-1} in the $\ket{0}$, $\ket{1}$ basis are real, we have
\begin{equation}
KH_{v}K=-H_{v}.
\end{equation}
Thus one can calculate in the $\ket{0}$, $\ket{1}$ basis 
\begin{equation}
\left(-i\sigma_{y}\right)KH_{v}K\left(i\sigma_{y}\right)=H_{v}.
\end{equation}
If the system is initially in the ground the state $\ket{1}$ and
solution is shown as Eq. \eqref{eq:psi-t}, then the time reversal
operation will transform the initial state from $\ket{1}$ to $-i\sigma_{y}K\ket{1}=e^{i\pi}\ket{0}$
and the solution for the system initially prepared in the excited
state $e^{i\pi}\ket{0}$ would be 
\begin{equation}
\ket{\widetilde{\psi}(t)}=-i\sigma_{y}K\ket{\psi(t)}=-C_{1}^{*}(t)\ket{0}+C_{0}^{*}(t)\ket{1}.\label{eq:psi-til}
\end{equation}
Plugging Eqs. (\ref{eq:psi-t}, \ref{eq:psi-til}) into Eq. \eqref{eq:Ig-def},
we find $I_{v,\,\ket{1}}=I_{v,\,\ket{0}}$. By similar arguments,
we have $I_{-v,\,\ket{1}}=I_{-v,\,\ket{0}}$.$\blacksquare$ 

The equality of the CFIs for the four cases follows from the same
argument as above. In what follows, we shall restrict the system in
the condition of case $(1)$. Once we know the Fisher information
or optimal measurements for this case, those for the rest three cases
are also known by the virtue of Eqs. (\ref{eq:psi-hat}, \ref{eq:psi-til}).

\section{The classical Fisher information based on Landau-Zener formula\label{sec:states}}

In this section, following Zener's \cite{zener_non-adiabatic_1932}
derivations, we express the general solution to the evolution of system
evolving under LZ Hamiltonian in terms of parabolic cylinder function
$D_{\nu}(z)$ and then find the state at $t=T$ and the classical
Fisher informations(CFIs) measured in the $\sigma_{z}$ basis.

The Schödinger equation for a system evolving under the LZ Hamiltonian
\eqref{eq:H-1} in the $\sigma_{z}$ basis is written as 
\begin{equation}
\left\{ \begin{array}{c}
i\dot{C}_{0}(t)=\frac{vt}{2}C_{0}(t)+\frac{\Delta}{2}C_{1}(t)\\
i\dot{C}_{1}(t)=\frac{\Delta}{2}C_{0}(t)-\frac{vt}{2}C_{1}(t)
\end{array}\right..\label{eq:govCs}
\end{equation}
Assuming $v>0$ and the system is initially in the ground state $\ket{1}$
at initial time $t_{i}=-T_{0}$, i.e., 
\begin{equation}
\begin{array}{cc}
C_{0}(-T_{0})=0, & C_{1}(-T_{0})=1,\end{array}\label{eq:intial0}
\end{equation}
 Eliminating $C_{1}(t)$ in Eq. \eqref{eq:govCs}, we have a second
order differential equation with respect to $C_{0}(t)$, i.e.,

\begin{equation}
\ddot{C}_{0}(t)+\left(\dfrac{\Delta^{2}+v^{2}t^{2}}{4}+i\dfrac{v}{2}\right)C_{0}(t)=0.\label{eq:2nd_C1}
\end{equation}
The change of variables $z=e^{\frac{\pi i}{4}}\sqrt{v}t$ transforms
Eq.\eqref{eq:2nd_C1} into the Weber equation 
\begin{equation}
C_{0}^{\prime\prime}(z)+\left[\nu+\dfrac{1}{2}-\dfrac{z^{2}}{4}\right]C_{0}(z)=0,
\end{equation}
where $\nu\equiv-i\gamma$ and $\gamma=\frac{\Delta^{2}}{4v}$. The
linearly independent solutions of the Weber equation is the parabolic
cylinder functions $D_{\nu}(z)$ and $D_{-\nu-1}(iz)$(sect. 16.5
\cite{whittaker_course_1996}; sect 6.12\cite{wang_special_1989}).
Let us further assume $T_{0}$ is very large, then at $t=-T_{0}$,
$z=\sqrt{v}T_{0}e^{-\frac{3\pi i}{4}}$ $\equiv R_{0}e^{-\frac{3\pi i}{4}}$
and $D_{-\nu-1}(iz)$ becomes $D_{-\nu-1}(R_{0}e^{-\frac{\pi i}{4}})$,
where $R_{0}\equiv\sqrt{v}T_{0}$ is a the real dimensionless large
quantity. According to the asymptotic expansion of $D_{n}(z)$ when
$\big|z\big|\to\infty$ for $\big|\mathrm{arg}z\big|<\frac{3\pi}{4}$
(sect. 16.5 \cite{whittaker_course_1996}; sect 6.12\cite{wang_special_1989})

\begin{equation}
D_{n}(z)\sim e^{-\frac{z^{2}}{4}}z^{n}\left\{ 1-\dfrac{n(n-1)}{2z^{2}}+\dfrac{n(n-1)(n-2)(n-3)}{2\cdot4z^{4}}-\cdots\right\} ,\label{eq:D_Asymp}
\end{equation}
we have at $t=-T_{0}$

\begin{equation}
\lim_{R_{0}\to\infty}D_{-\nu-1}(R_{0}e^{-\frac{\pi i}{4}})=\lim_{R_{0}\to\infty}e^{\frac{iR_{0}^{2}}{4}}e^{\frac{i\pi(\nu+1)}{4}}R_{0}^{-\nu-1}=0.\label{eq:D-nu-1}
\end{equation}
Eq. \eqref{eq:D-nu-1} implies that we can assume 

\begin{equation}
C_{0}(z)=AD_{-\nu-1}(iz),\label{eq:general_C1}
\end{equation}
to satisfy the boundary of the first equation of Eq. \eqref{eq:intial0}.
By noting the recurrence relations of $D_{\nu}(z)$, 
\begin{equation}
D_{\nu}^{\prime}(z)+\dfrac{z}{2}D_{\nu}(z)-\nu D_{\nu-1}(z)=0,
\end{equation}
we obtain that
\begin{equation}
D_{-\nu-1}^{\prime}(iz)=\dfrac{z}{2}D_{-\nu-1}(iz)-i(\nu+1)D_{-\nu-2}(iz).\label{eq:D_prime}
\end{equation}
Substitution of Eq. \eqref{eq:general_C1} and Eq. \eqref{eq:D_prime}
into the first equation of Eq. \eqref{eq:govCs} yields 
\begin{equation}
C_{1}(z)=-\dfrac{2A\sqrt{v}}{\Delta}e^{-\frac{3\pi i}{4}}\left[izD_{-\nu-1}\left(iz\right)+\left(\nu+1\right)D_{-\nu-2}\left(iz\right)\right].\label{eq:general_C2}
\end{equation}
From Eq. \eqref{eq:D_Asymp}, we can easily find at $t=-T_{0}$, $z=R_{0}e^{-\frac{3\pi i}{4}}\to\infty e^{-\frac{3\pi i}{4}}$,
\begin{equation}
\lim_{z\to\infty e^{-\frac{3\pi i}{4}}}zD_{-\nu-1}(iz)=\lim_{R_{0}\to\infty}R_{0}e^{-\frac{3\pi i}{4}}D_{-\nu-1}(R_{0}e^{-\frac{\pi i}{4}})=\lim_{R_{0}\to\infty}e^{\frac{iR_{0}^{2}}{4}-\frac{i\pi}{2}}R_{0}^{i\gamma}e^{\frac{i\pi\nu}{4}}=e^{\frac{\pi\gamma}{4}}e^{i(\phi_{0}(T_{0})-\frac{3\pi}{4})},\label{eq:zD-nu-1}
\end{equation}

\begin{equation}
\lim_{z\to\infty e^{-\frac{3\pi i}{4}}}D_{-\nu-2}(iz)=\lim_{R_{0}\to\infty}D_{-\nu-2}(R_{0}e^{-\frac{\pi i}{4}})=\lim_{R_{0}\to\infty}e^{\frac{iR_{0}^{2}}{4}}e^{\frac{i\pi(\nu+2)}{4}}R_{0}^{-\nu-2}=0,\label{eq:D-nu-2}
\end{equation}
where we define $\phi_{0}(T_{0})\equiv\frac{R_{0}^{2}+\pi}{4}+\gamma\ln R_{0}$
and the real dimensionless quantity $\gamma\equiv\frac{\Delta^{2}}{4v}$.
From Eqs. (\ref{eq:general_C2}, \ref{eq:zD-nu-1}, \ref{eq:D-nu-2}),
we find
\begin{equation}
C_{1}\left\{ z\to\infty\exp(-3\pi i\big/4)\right\} =\dfrac{2A\sqrt{v}}{\Delta}e^{\frac{\pi\gamma}{4}}e^{i\phi_{0}(T_{0})}.\label{eq:C2zinf}
\end{equation}
Comparing Eq. \eqref{eq:C2zinf} with the second equation of Eq. \eqref{eq:intial0},
we determine $A$ as

\begin{equation}
A=\dfrac{\Delta}{2\sqrt{v}}e^{-\frac{\pi\gamma}{4}}=\sqrt{\gamma}e^{-\frac{\pi\gamma}{4}}e^{-i\phi_{0}},\label{eq:coeff-A}
\end{equation}
where in the calculations the phases cancel with each other, leaving
$A$ a real number, which is the reason we set the phase $\phi_{1}$
in the initial condition as Eq. \eqref{eq:intial0}. Therefore

\begin{equation}
C_{0}(z)=\sqrt{\gamma}e^{-\frac{\pi\gamma}{4}}e^{-i\phi_{0}}D_{-\nu-1}(iz).\label{eq:C0z}
\end{equation}
Substitution of Eqs. (\ref{eq:general_C1}, \ref{eq:D_prime}) into
the first equation of Eq. \eqref{eq:govCs} yields

\begin{equation}
C_{1}(z)=e^{-\frac{\pi\gamma}{4}}e^{-i\phi_{0}+\pi i/4}\left[izD_{-\nu-1}\left(iz\right)+\left(\nu+1\right)D_{-\nu-2}\left(iz\right)\right].\label{eq:C1z}
\end{equation}
 For $\frac{\pi}{4}<\text{arg}(z)<\frac{5\pi}{4}$, as $\big|z\big|\to\infty$,
by using the identity (sect. 16.5 \cite{whittaker_course_1996}; sect
6.12\cite{wang_special_1989})
\begin{equation}
D_{n}(z)=e^{n\pi i}D_{n}(-z)+\dfrac{\sqrt{2\pi}}{\Gamma(-n)}e^{\frac{1}{2}(n+1)\pi i}D_{-n-1}(-iz)\label{eq:Dn-Dn}
\end{equation}
and Eq. \eqref{eq:D_Asymp}, we can readily obtain the asymptotic
expansion 
\begin{equation}
D_{n}(z)=e^{-\frac{z^{2}}{4}}z^{n}\left\{ 1+O\left(\dfrac{1}{z^{2}}\right)\right\} -\dfrac{\sqrt{2\pi}}{\Gamma\left(-n\right)}e^{n\pi i}e^{\frac{z^{2}}{4}}z^{-n-1}\left\{ 1+O\left(\dfrac{1}{z^{2}}\right)\right\} .\label{eq:Dinf1}
\end{equation}
We are concerned with the state at time $t=T$ which is far away from
the avoided level crossing time $t=0$. Therefore at $t=T$, $z=Re^{\frac{\pi i}{4}}$,
$iz=\sqrt{v}Te^{\frac{3\pi i}{4}}\equiv Re^{\frac{3\pi i}{4}}$, $D_{-\nu-1}(iz)=D_{-\nu-1}(Re^{\frac{3\pi i}{4}})$
and $D_{-\nu-2}(iz)=D_{-\nu-2}(Re^{\frac{3\pi i}{4}})$, using Eq.\eqref{eq:Dinf1},
we have

\begin{equation}
\lim_{R\to\infty}D_{-\nu-1}(Re^{\frac{3\pi i}{4}})=\lim_{R\to\infty}\left\{ e^{-\frac{3\pi i}{4}(\nu+1)}e^{\frac{iR^{2}}{4}}R^{-\nu-1}+\dfrac{\sqrt{2\pi}}{\Gamma(1+\nu)}e^{-\frac{\nu\pi i}{4}}e^{-\frac{iR^{2}}{4}}R^{\nu}\right\} =\dfrac{\sqrt{2\pi}}{\Gamma(1+\nu)}e^{-\frac{\pi\gamma}{4}}e^{-i(\phi-\frac{\pi}{4})},\label{eq:D-nu-1T}
\end{equation}

\begin{eqnarray}
\lim_{R\to\infty}RD_{-\nu-1}(Re^{\frac{3\pi i}{4}}) & = & \lim_{R\to\infty}\left\{ e^{-\frac{3\pi i}{4}(\nu+1)}e^{\frac{iR^{2}}{4}}R^{-\nu}\right.\left.+\dfrac{\sqrt{2\pi}}{\Gamma(\nu+1)}e^{-\frac{\nu\pi i}{4}}e^{-\frac{iR^{2}}{4}}R^{\nu+1}\right\} \nonumber \\
 & = & e^{-\frac{3\pi\gamma}{4}}e^{i(\phi-\pi)}+\dfrac{\sqrt{2\pi}}{\Gamma(1+\nu)}e^{-\frac{\pi a}{2}}Re^{-i(\phi-\frac{\pi}{4})},
\end{eqnarray}

\begin{equation}
\lim_{R\to\infty}D_{-\nu-2}(Re^{\frac{3\pi i}{4}})=\lim_{R\to\infty}\left\{ e^{-\frac{3\pi i}{4}(\nu+2)}e^{\frac{iR^{2}}{4}}R^{-\nu-2}\right.\left.+\dfrac{\sqrt{2\pi}}{\Gamma(2+\nu)}e^{-\frac{(\nu+1)\pi i}{4}}e^{-\frac{iR^{2}}{4}}R^{\nu+1}\right\} =\dfrac{\sqrt{2\pi}}{\Gamma(2+\nu)}e^{-\frac{\pi\gamma}{4}}Re^{-i\phi}.\label{eq:D-nu-2-T}
\end{equation}
where we define $\phi(T)\equiv\frac{R^{2}+\pi}{4}+\gamma\ln R$. According
to Eqs. \eqref{eq:C0z} and (\ref{eq:D-nu-1T}-\ref{eq:D-nu-2-T}),
we arrive at
\begin{equation}
C_{0}(T)=e^{-\frac{\pi\gamma}{2}}\dfrac{\sqrt{2\pi\gamma}}{\Gamma(1-i\gamma)}e^{-i(\phi(T)+\phi_{0}(T_{0})-\frac{\pi}{4})},\label{eq:C0Tbar}
\end{equation}

\begin{equation}
C_{1}(T)=e^{-\pi\gamma}e^{i(\phi(T)-\phi_{0}(T_{0}))}.\label{eq:C1Tbar}
\end{equation}
Multiplying Eqs. (\ref{eq:C0Tbar}, \ref{eq:C1Tbar}) by an overall
phase to eliminate the phase factor in $C_{1}(T)$, we obtain the
state at the time $t=T$ far away from the transition region as

\begin{equation}
C_{0}(T)=e^{-\frac{\pi\gamma}{2}}\dfrac{\sqrt{2\pi\gamma}}{\Gamma(1-i\gamma)}e^{-i(2\phi(T)-\frac{\pi}{4})},\label{eq:C0T}
\end{equation}

\begin{equation}
C_{1}(T)=e^{-\pi\gamma}.\label{eq:C1T}
\end{equation}
From Eqs. \eqref{eq:C0T} and \eqref{eq:C1T}, we easily obtain
\begin{equation}
P_{0}(T)\equiv\big|C_{1}(T)\big|^{2}=1-e^{-2\pi\gamma},\label{eq:C1TSQ}
\end{equation}

\begin{equation}
P_{1}(T)\equiv\big|C_{2}(T)\big|^{2}=e^{-2\pi\gamma}.\label{eq:C2TSQ-1}
\end{equation}
where we have used $\left[\Gamma(z)\right]^{*}=\Gamma(z^{*})$ and
$\Gamma\left(1+i\gamma\right)\Gamma(1-i\gamma)=\frac{\pi\gamma}{\sinh(\pi\gamma)}$
(\cite{whittaker_course_1996,wang_special_1989}). The transition
probabilities $P_{0}(T)$, $P_{1}(T)$ in Eqs. (\ref{eq:C1TSQ}, \ref{eq:C2TSQ-1})
are exactly the same as the results in Zener's paper\cite{zener_non-adiabatic_1932}. 

\section{expressions for the QFIs at $t=T$ for the case of a single Landau-Zener
transition\label{sec:Exact-QFIs}}

In this section, we present a more accurate asymptotic for the QFI
of estimating $\Delta$ and show how the time scaling of the QFI when
the system is initially in a superposition state. 

Asymptotically expanding Eqs. (\ref{eq:C0z}, \ref{eq:C1z}) at $t_{i}=-T_{0}$
gives
\begin{equation}
C_{0}(-T_{0})=\sqrt{\gamma}\left\{ 0+\dfrac{1}{R_{0}}-i\dfrac{(\nu+1)(\nu+2)}{2R_{0}^{3}}-\dfrac{(\nu+1)(\nu+2)(\nu+3)(\nu+4)}{8R_{0}^{5}}+\cdots\right\} ,\label{eq:AsymC0-T}
\end{equation}
\begin{equation}
C_{1}(-T_{0})=1-i\dfrac{\nu(\nu+1)}{2R_{0}^{2}}+\dfrac{\left(\nu+1\right)(\nu+2)(\nu+3)}{2R_{0}^{4}}+\cdots.\label{eq:AsymC1-T}
\end{equation}
From Eq. (\ref{eq:AsymC0-T}, \ref{eq:AsymC1-T}), it is easily identified
that in order to make the derivations in Sec. \ref{sec:states} accurate,
we should require $\sqrt{\gamma}/R_{0}\ll1$ and $1/R_{0}\ll1$, from
which we obtain 
\begin{equation}
T_{0}\gg\tau\equiv\max\{\dfrac{\Delta}{2v},\,\dfrac{1}{\sqrt{v}}\}.\label{eq:tau}
\end{equation}
Asymptotically expanding Eqs. (\ref{eq:C0z}, \ref{eq:C1z}) at $t=T$
with Eqs. (\ref{eq:D_Asymp}, \ref{eq:Dn-Dn}) shows that 
\begin{equation}
T\gg\tau\label{eq:Tggtau}
\end{equation}
 is also sufficient to make the asymptotic results Eqs. (\ref{eq:C0T},
\ref{eq:C1T}) sufficient accurate. 

Strictly speaking the solutions Eqs. (\ref{eq:C0z}, \ref{eq:C1z})
to the time-dependent Schödinger Eq. \eqref{eq:govCs} is asymptotic
rather than exact since their asymptotic expansions at $t_{i}=-T_{0}$
(\ref{eq:AsymC0-T}, \ref{eq:AsymC1-T}) satisfy the initial condition
\eqref{eq:intial0} asymptotically rather than exactly. So one may
construct the exact solution by considering the small corrections
in Eqs. (\ref{eq:AsymC0-T}, \ref{eq:AsymC1-T}), i.e. $C_{i}(t)=\sum_{n=0}^{\infty}C_{i}^{(n)}(t)(i=1,\,2)$,
where $C_{i}^{(0)}$ corresponding the solution Eqs. (\ref{eq:C0z},
\ref{eq:C1z}). One can safely ignore the higher order small corrections
and only keep the zeroth order to find the transition probabilities
as Zener did and the relative phase at $t=T$. However, since the
quantum Fisher information involves the derivative of the amplitudes
$C_{i}$'s with respect to the estimation parameter $g$ as one can
see from Eq. \eqref{eq:Ig-def}, the derivatives of these small high
order corrections are not necessarily small and hence may contribute
significant the quantum Fisher information. In fact, we show in our
subsequent paper that the time scaling of QFI of $v$ at $t=0$ is
due to the first order rather than the zeroth order while for the
scaling of QFI of $v$ at $t=T$, we only need to consider to zeroth
order since it gives rise to the highest time scaling $T^{4}$. We
also show for QFIs of $\Delta$ at $t=0$ and $t=T$, only the zeroth
order terms contribute to the QFIs and the contributions from high
order correction can be neglected as long as $T$ is large. These
observations justify that in the main text we only use the zero-order
solution (\ref{eq:C0T}, \ref{eq:C1T}) to find the time scaling of
the QFIs of $\Delta$ and $v$ at $t=T$.

\subsection{Improved asymptotic expression for the QFI of estimating $\Delta$}

In main text, we show that the QFI scales as $\ln T$ for estimating
$\Delta$ and $T^{4}$ for estimating $v$ with a prefactor given
by the product of the two transition probabilities $P_{0}$ and $P_{1}$.
In either a diabatic or adiabatic transition, one of the transition
probabilities is very small, which will also make their product small.
But for estimating $v$, the quickly increasing $T^{4}$ term will
compensate for the smallness of the prefactor, which make the $1^{st}$
asymptotic for estimating $v$ in the main text still be dominantly
large as long as Eqs. (\ref{eq:tau}, \ref{eq:Tggtau}) is satisfied.

However, this is not the case for estimating $\Delta$. In either
a diabatic or adiabatic transition, the small prefactor $P_{0}P_{1}$
conspires with the slowness of the $\ln T$ scaling, which makes the
$1^{st}$ asymptotic for estimating $\Delta$ in the main text no
longer large enough to be the leading term, as we will see in the
end of this subsection. 

Therefore, a more precise asymptotic expression for QFI of $\Delta$
is necessary particularly for the diabatic or adiabatic case. Starting
from Eqs. (\ref{eq:C0T}, \ref{eq:C1T}), we will calculate asymptotic
expression for the QFI of $\Delta$ by keeping all the terms in the
calculation rather than only keeping the the highest order terms of
$T$. First let us rewrite Eq. \eqref{eq:Ig-def} as 
\begin{equation}
I_{g}(t)=4\left[\sum_{i=0}^{1}\Big|C_{i}(t)\Big|^{2}\Big|\partial_{g}\ln C_{i}(t)\Big|^{2}\right.\left.-\Big|\sum_{i=0}^{1}\Big|C_{i}(t)\Big|^{2}\partial_{g}\ln C_{i}(t)\Big|^{2}\right].\label{eq:QFI_Computation}
\end{equation}
Taking the logarithm on both sides of Eqs. (\ref{eq:C0T}, \ref{eq:C1T}),
we obtain

\begin{equation}
\ln C_{0}(T)=\dfrac{1}{2}\ln\gamma-\dfrac{\pi\gamma}{2}+\dfrac{1}{2}\ln2+\dfrac{1}{2}\ln\pi-\ln\Gamma\left(1-i\gamma\right)-2i\phi+i\frac{\pi}{4},\label{eq:lnC1T}
\end{equation}

\begin{equation}
\ln C_{1}(T)=-\pi\gamma.\label{eq:lnC2T}
\end{equation}
Differentiating with respect to $\Delta$ in both sides of Eq. (\ref{eq:lnC1T},
\ref{eq:lnC2T}), by noticing $\partial_{\Delta}\gamma=\frac{2\gamma}{\Delta}$
and $\partial_{\Delta}\phi=\dfrac{\gamma}{\Delta}\ln\left(vT^{2}\right)$,
we arrive at

\begin{equation}
\partial_{\Delta}\ln C_{0}(T)=\dfrac{1-\pi\gamma}{\Delta}+i\dfrac{2\gamma}{\Delta}\varphi(1-i\gamma),\label{eq:dlnC1TdDel}
\end{equation}
\begin{equation}
\partial_{\Delta}\ln C_{1}(T)=-\dfrac{2\pi\gamma}{\Delta}+i\dfrac{2\gamma}{\Delta}\ln\left(vT^{2}\right),\label{eq:dlnC2TdDel}
\end{equation}
where $\varphi(1-i\gamma)$ is the digamma function defined in Eq.
\eqref{eq:digamma}. Substitution of Eq. (\ref{eq:varphi1}, \ref{eq:eta1_closed})
into Eq. \eqref{eq:dlnC1TdDel} yields

\begin{equation}
\partial_{\Delta}\ln C_{0}(T)=\dfrac{1-\pi\gamma+2\gamma\eta_{1}(\gamma)}{\Delta}+i\dfrac{2\gamma}{\Delta}\theta_{1}(\gamma)=\dfrac{\pi\gamma\left[\coth\left(\pi\gamma\right)-1\right]}{\Delta}+i\frac{2\gamma}{\Delta}\theta_{1}(\gamma),\label{eq:dlnC1TdDelCompact}
\end{equation}
where $\theta_{1}(\gamma)$ is defined as Eq. \eqref{eq:theta-nu}.
Plugging Eqs. (\ref{eq:C1TSQ}, \ref{eq:C2TSQ-1}, \ref{eq:dlnC1TdDelCompact},
\ref{eq:dlnC2TdDel}) into Eq. \eqref{eq:QFI_Computation} yields
\begin{eqnarray}
I_{\Delta,\,\ket{1}}(T) & = & \dfrac{16\gamma^{2}e^{-2\pi\gamma}}{\Delta^{2}\left(1-e^{-2\pi\gamma}\right)}\left\{ \left(1-e^{-2\pi\gamma}\right)^{2}\left[\ln\left(vT^{2}\right)\right]^{2}\right.-2\left(1-e^{-2\pi\gamma}\right)\theta_{1}(\gamma)\ln\left(vT^{2}\right)\nonumber \\
 &  & \left.+\left(\pi^{2}+\theta_{1}^{2}(\gamma)-2e^{-2\pi\gamma}\theta_{1}^{2}(\gamma)+\theta_{1}^{2}(\gamma)e^{-4\pi\gamma}\right)\right\} .\label{eq:IDelT-temp}
\end{eqnarray}

\begin{figure}
\begin{centering}
\includegraphics[scale=0.58]{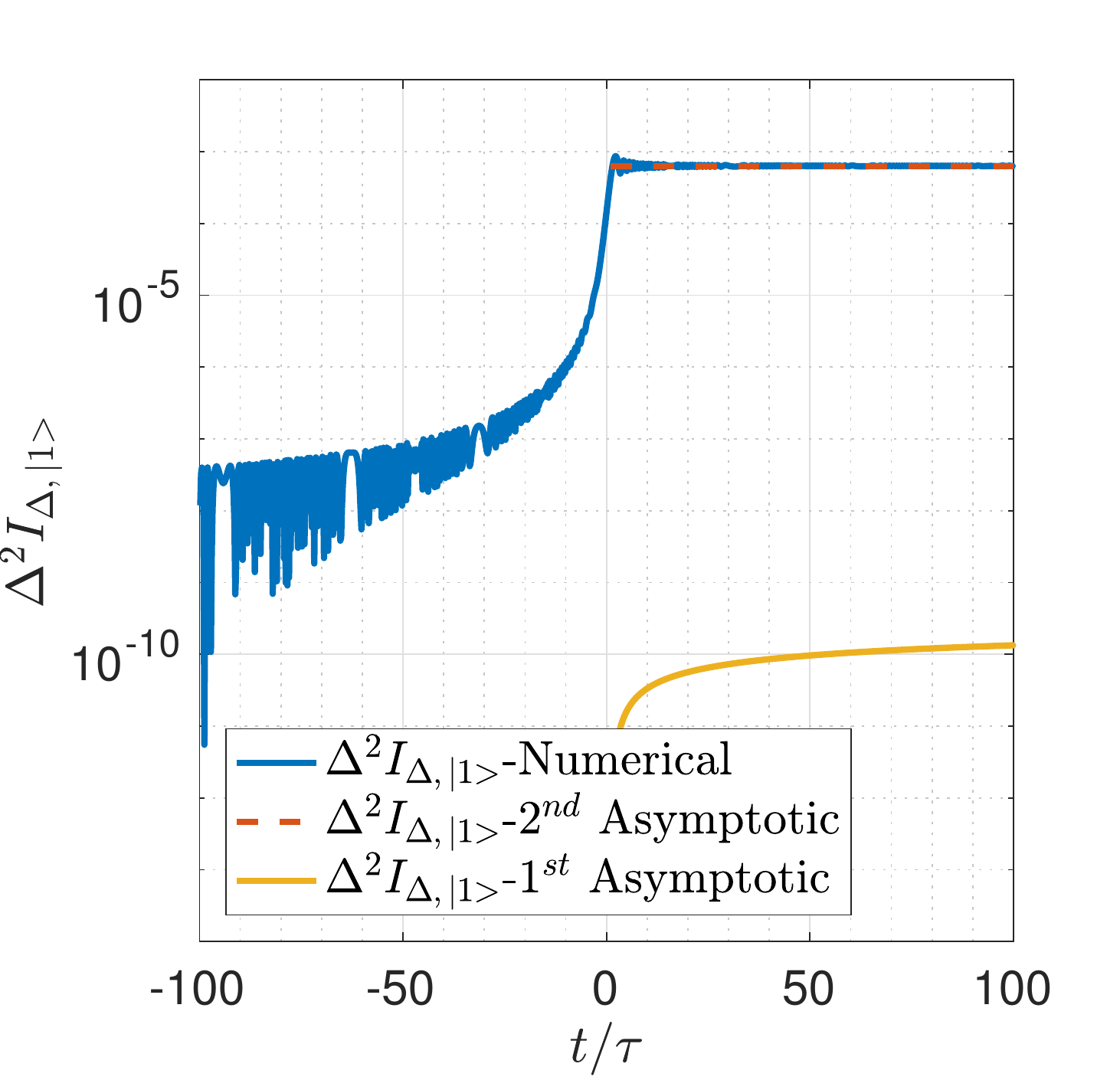}
\par\end{centering}
\caption{Comparison of the analytic results of estimating $\Delta$ with the
numerical and asymptotic results mentioned in the main text. We consider
the diabatic limit $v\gg\Delta$: $v=1$, $\Delta=0.01$, $\tau=1$,
$T_{0}=100\tau$. One see that the $1^{st}$ asymptotic expression
in the main text deviates from the numerical result significantly
while the $2^{nd}$ asymptotic expression \eqref{eq:IDelT} agrees
with the numerical result very well. \label{fig:ana-del}}
\end{figure}

In view of Eqs. (\ref{eq:C1TSQ}, \ref{eq:C2TSQ-1}), Eq. \eqref{eq:IDelT-temp}
can be rewritten as 
\begin{equation}
I_{\Delta,\,\ket{1}}(T)=\frac{\Delta^{2}}{v^{2}}P_{0}P_{1}\left[\ln\left(vT^{2}\right)\right]^{2}-\frac{2\Delta^{2}}{v^{2}}P_{1}\theta_{1}(\gamma)\ln\left(vT^{2}\right)+\frac{\Delta^{2}P_{1}}{v^{2}P_{0}}\left[\left(\pi^{2}+\theta_{1}^{2}(\gamma)-2P_{1}\theta_{1}^{2}(\gamma)+\theta_{1}^{2}(\gamma)P_{1}^{2}\right)\right].\label{eq:IDelT}
\end{equation}
The first term in our $2^{nd}$ asymptotic expression Eq. \eqref{eq:IDelT}
is just the $1^{st}$ asymptotic expression discussed in the main
text and in the diabatic or adiabatic limit the remaining terms will
become at least comparable with (even much larger than) the first
term. Fig. \ref{fig:ana-del} illustrates this in the diabatic limit
where the $1^{st}$ asymptotic result in the main text is not longer
dominant and the $2^{nd}$ asymptotic expression \eqref{eq:IDelT}
is needed for an accurate predication.

\subsection{Initially prepared in a superposition of states $\ket{0}$ and $\ket{1}$\label{sec:superpositionLZ}}

In this section we will analyze how the QFIs at $t=T$ scale as $T$
asymptotically when $T$ is large ($T\to\infty$) if we initially
prepare the state in a superposition state of $\ket{0}$ and $\ket{1}$.
Because the Schödinger equation Eq. \eqref{eq:govCs} is linear with
respect to $C_{i}$'s, the state for a system initially prepared in
a general state 
\begin{equation}
\Psi(-T)=\cos\left(\frac{\alpha}{2}\right)\ket{0}+e^{i\beta}\sin\left(\frac{\alpha}{2}\right)\ket{1}\label{eq:supPsi}
\end{equation}
is simply the linear superposition of the solutions corresponding
to the system initially prepared in $\ket{1}$ and $\ket{0}$, respectively.
Denoting the solution corresponding to initial condition \eqref{eq:supPsi}
as 
\begin{equation}
\ket{\check{\psi}(t)}=\mathcal{C}_{0}(t)\ket{0}+\mathcal{C}_{1}(t)\ket{1},
\end{equation}
then according to Eq. \eqref{eq:psi-til}, we can immediately write
the solution at $t=T$ as
\begin{equation}
\mathcal{C}_{0}(T)=\cos\left(\frac{\alpha}{2}\right)C_{0}(T)+i\sin\left(\frac{\alpha}{2}\right)e^{i\beta}\left[C_{1}(T)\right]^{*},\label{eq:C1supT}
\end{equation}
\begin{equation}
\mathcal{C}_{1}(T)=\cos\left(\frac{\alpha}{2}\right)C_{1}(T)-i\sin\left(\frac{\alpha}{2}\right)e^{i\beta}\left[C_{0}(T)\right]^{*}.\label{eq:C2supT}
\end{equation}

\begin{figure}
\begin{centering}
\includegraphics[scale=0.58]{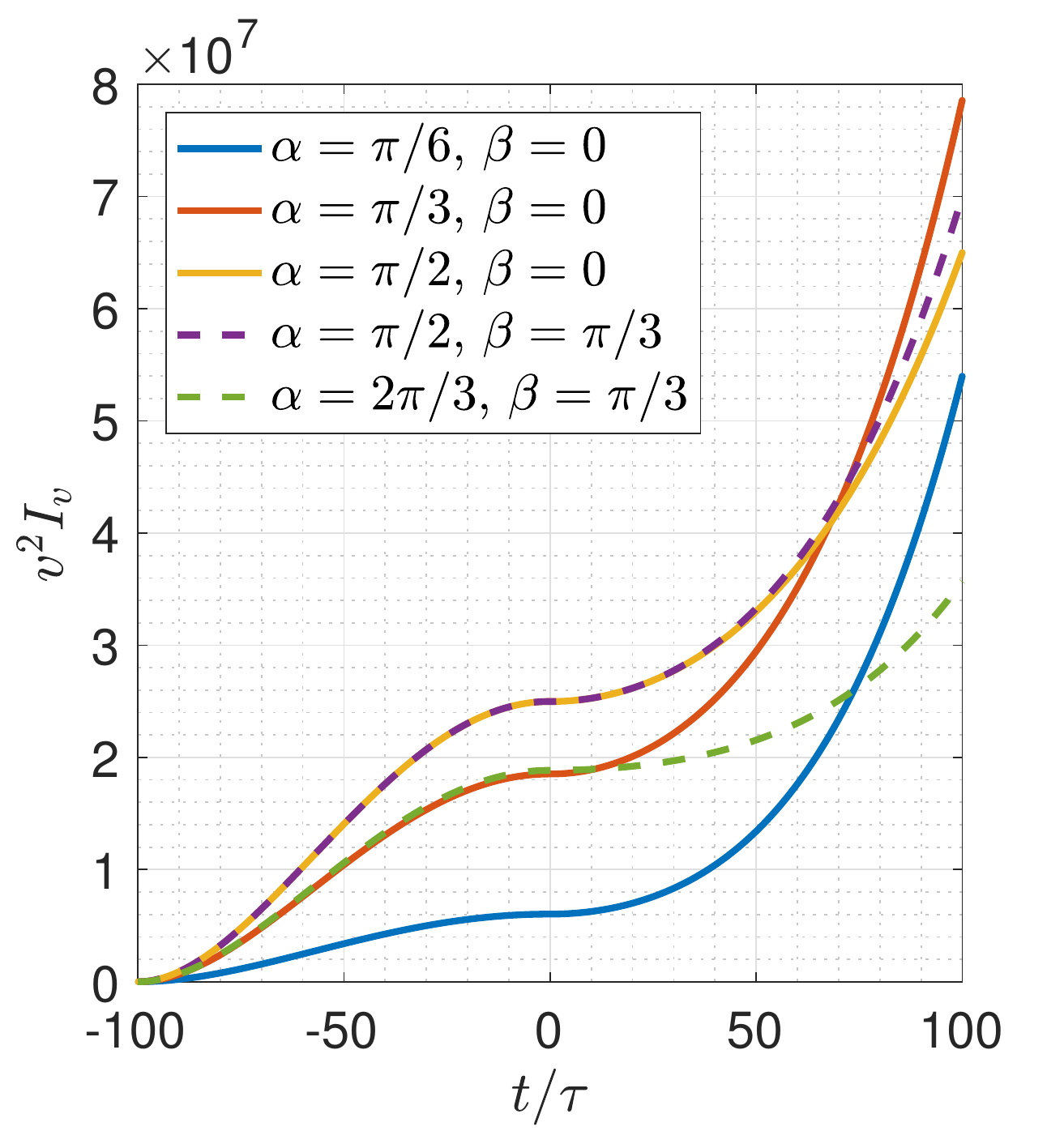}
\par\end{centering}
\caption{The QFIs of estimating $v$ by preparing the initially state in superpositions
of $\ket{0}$ and $\ket{1}$. The basic parameter configuration is
as follows: $v=1$, $\Delta=1$, $\tau=1$, $T_{0}=100\tau$. \label{fig:sup-v}}
\end{figure}

Given the results
\begin{equation}
\partial_{g}C_{0}(T)\sim-2iC_{0}(T)\partial_{g}\phi,
\end{equation}

\begin{equation}
\partial_{g}C_{1}(T)\sim\text{constant},
\end{equation}
\begin{equation}
\partial_{g}\mathcal{C}_{0}(T)\sim-2i\cos\left(\frac{\alpha}{2}\right)C_{0}(T)\partial_{g}\phi,\label{eq:dgdC1sup}
\end{equation}
 
\begin{equation}
\partial_{g}\mathcal{C}_{1}(T)\sim2\sin\left(\frac{\alpha}{2}\right)e^{i\beta}\left[C_{0}(T)\right]^{*}\partial_{g}\phi,\label{eq:dgdC2sup}
\end{equation}
Substitution of Eqs. (\ref{eq:C1supT}, \ref{eq:C2supT}, \ref{eq:dgdC1sup},
\ref{eq:dgdC2sup}) into Eq. \eqref{eq:Ig-def}, we can also find
in general $I_{\Delta,\,\ket{\Psi}}$ scales as $(\ln T)^{2}$ and
$I_{v,\,\ket{\Psi}}$ scales as $T^{4}$. The QFIs of estimating $v$
for different initial superposition states are plotted in Fig. \ref{fig:sup-v}.

\section{The QFI for the case of periodic Landau-Zener transitions}

In order to obtain an analytic expression of the QFI for the case
of periodic LZ transition, we set in the periodic LZ Hamiltonian $\epsilon_{0}=0$,
i.e., $H=\frac{1}{2}A\cos\left(\omega t\right)\sigma_{z}+\frac{1}{2}\Delta\sigma_{x}$.
Applying the transformation $\ket{\bar{\psi}}=e^{i\sigma_{y}\frac{\pi}{4}}\ket{\psi}$,
we obtain $i\partial_{t}\ket{\bar{\psi}}=H_{E}\ket{\bar{\psi}}$,
where 
\begin{equation}
H_{E}=e^{i\sigma_{y}\frac{\pi}{4}}He^{-i\sigma_{y}\frac{\pi}{4}}=\frac{\Delta}{2}\sigma_{z}-\frac{A}{2}\cos\left(\omega t\right)\sigma_{x}.
\end{equation}
Note that $H_{E}$ is the Hamiltonian of a two level atom driven by
a linear polarized monochromatic laser field. In the interaction picture,
where the Hamiltonian transforms into 
\begin{equation}
\mathcal{H}_{E}=e^{i\frac{\Delta}{2}\sigma_{z}t}\left[-\frac{A}{2}\cos\left(\omega t\right)\sigma_{x}\right]e^{-i\frac{\Delta}{2}\sigma_{z}t}=-\frac{A}{2}\cos(\omega t)\left[\sigma_{x}\cos(\Delta t)-\sigma_{y}\sin(\Delta t)\right],
\end{equation}
the Schödinger equation becomes $i\partial_{t}\ket{\psi}_{I}=\tilde{H}_{E}\ket{\psi}_{I}$,
where $\ket{\psi}_{I}$ denotes the state in the interaction picture.
Upon writing $\ket{\psi(t)}_{I}=c_{0}(t)\ket{0}+c_{1}\ket{1}$ and
substituting it into the Schödinger equation in the interaction picture,
we obtain \begin{subequations}

\begin{eqnarray}
i\dot{c}_{0}(t) & = & -\dfrac{A}{4}\left[e^{i(\omega-\Delta)t}+e^{-i(\omega+\Delta)t}\right]c_{1}(t)\label{eq:c0}\\
i\dot{c}_{1}(t) & = & -\dfrac{A}{4}\left[e^{-i(\omega-\Delta)t}+e^{i(\omega+\Delta)t}\right]c_{0}(t)\label{eq:c1}
\end{eqnarray}
\end{subequations}Let us consider the weakly coupling and near resonance
case, where $A\ll\omega,\,\Delta$ and $|\omega-\Delta|\ll|\omega+\Delta|$
, then we can apply the rotating wave approximation , i.e., neglecting
the anti-rotating terms in Eqs. (\ref{eq:c0}, \ref{eq:c1}) and arrive
at 
\begin{eqnarray}
i\dot{c}_{0}(t) & = & -\dfrac{A}{4}e^{i(\omega-\Delta)t}c_{1}(t),\\
i\dot{c}_{1}(t) & = & -\dfrac{A}{4}e^{-i(\omega-\Delta)t}c_{0}(t),
\end{eqnarray}
from which one can easily identify that the Hamiltonian now is simplified
as
\begin{equation}
\mathcal{H}_{E}=-\dfrac{A}{4}\left[\begin{array}{cc}
0 & e^{i(\omega-\Delta)t}\\
e^{-i(\omega-\Delta)t} & 0
\end{array}\right]=-\dfrac{A}{4}\left\{ \cos\left[(\omega-\Delta)t\right]\sigma_{x}-\sin\left[(\omega-\Delta)t\right]\sigma_{y}\right\} =-\dfrac{A}{4}e^{i\sigma_{z}\frac{\omega-\Delta}{2}t}\sigma_{x}e^{-i\sigma_{z}\frac{\omega-\Delta}{2}t}.\label{eq:nmr}
\end{equation}
Eq. \eqref{eq:nmr} is a well-known Hamiltonian first considered by
Rabi in nuclear magnetic resonance. From Eq. \eqref{eq:nmr}, one
can easily transform $\mathcal{H}_{E}$ to a static Hamiltonian $-\dfrac{A}{4}\left[\left(\frac{\omega-\Delta}{2}\right)\sigma_{z}+\sigma_{x}\right]$
and therefore find the corresponding analytical solution. Now the
state $\ket{\psi(t)}$ corresponding to the original Hamiltonian $H$
and the state $\ket{\psi(t)}_{E}$ corresponding to the Hamiltonian
$\mathcal{H}_{E}$ are connected by the unitary transformation
\begin{equation}
\ket{\psi(t)}_{E}=e^{i\frac{\Delta}{2}\sigma_{z}t}e^{i\sigma_{y}\frac{\pi}{4}}\ket{\psi(t)}\equiv U_{E}\ket{\psi(t)}.\label{eq:psi-E}
\end{equation}
Since the the transformation $U_{E}$ does not depend on the estimation
parameter $\omega$, plugging Eq. \eqref{eq:psi-E} into Eq. \eqref{eq:Ig-def},
we see that the the QFI in the transformed frame is identical to that
in the original lab frame. On the the other hand, starting from $t=0$,
the maximum quantum Fisher information at $t=T$ over all possible
initial pure states associated with $\mathcal{H}_{E}$ is, according
to \cite{pang_quantum_2016},
\begin{equation}
\max I_{\omega}(T)=\dfrac{A^{2}T^{2}}{A^{2}+4\left(\Delta-\omega\right)^{2}}-\dfrac{4A^{2}T\sin\left(T\sqrt{A^{2}+4(\Delta-\omega)^{2}}\big/2\right)}{\left[A^{2}+4\left(\Delta-\omega\right)^{2}\right]^{3/2}}+\dfrac{8A^{2}\left[1-\cos\left(T\sqrt{A^{2}+4(\Delta-\omega)^{2}}\big/2\right)\right]}{\left[A^{2}+4\left(\Delta-\omega\right)^{2}\right]^{2}}.
\end{equation}
The corresponding initial state that gives rise to this maximum QFI
in general depends on the evolution duration $T$.

\section{the Optimal Level Crossing Hamiltonians and Measurements\label{sec:OptimalControlsMeas}}

In this section, we briefly introduce the expression for the optimal
quantum controls and measurements for the single (LZ) transition and
multiple periodic transitions.

The Optimal Control Hamiltonian(OCH) mentioned in the main text is

\begin{equation}
H_{c}\left(t\right)=\sum_{k}f_{k}\left(t\right)\ket{\psi_{k}\left(t\right)}\bra{\psi_{k}\left(t\right)}-H_{g}(t)+i\sum_{k}\ket{\partial_{t}\psi_{k}\left(t\right)}\bra{\psi_{k}\left(t\right)},\label{eq:HC_general}
\end{equation}
where $\ket{\psi_{k}\left(t\right)}$ is the $k$th eigenstate of
$\partial_{g}H_{g}$; $f_{k}(t)$ can be arbitrary taken in principle,
but is usually chosen to take the form which simplifies the OCH $H_{c}\left(t\right)$
significantly. Intuitively, the following Optimal Level Crossing Hamiltonian
(OLCH)
\begin{equation}
H_{LC}=h(t)\left\{ e^{i\left[\theta_{m}(t)-\theta_{n}(t)\right]}\ket{\psi_{n}(t)}\bra{\psi_{m}(t)}\right.\left.+e^{i\left[\theta_{n}(t)-\theta_{m}(t)\right]}\ket{\psi_{m}(t)}\bra{\psi_{n}(t)}\right\} ,\label{eq:Ha-general}
\end{equation}
 where $h(t)$ is proportional to a delta function peaked at the level
crossing time $t^{\prime}$ of the maximum of minimum energy levels
of $\partial_{g}H$, satisfying 
\begin{equation}
h(t)=(l+\frac{1}{2})\pi\delta(t-t^{\prime}),\label{eq:ht}
\end{equation}
where $l$ is an arbitrary integer and 
\begin{equation}
\theta_{k}(t)=\int_{t_{0}}^{t}f_{k}(t^{\prime})dt^{\prime},\label{eq:thetak}
\end{equation}
can transform the old state $\ket{\psi_{n}}$ (before level crossing)
to a new state $\ket{\psi_{m}}$ (after level crossing) and vice versa.
For a rigorous proof, see \cite{pang_quantum_2016}.

\subsection{A Single LZ Transition with optimal controls and measurements }

For the estimation of $\Delta$, taking $f_{\ket{+x}}=\dfrac{\Delta_{c}}{2}$
and $f_{\ket{-x}}=-\dfrac{\Delta_{c}}{2}$, the Optimal Control Hamiltonian
(OCH) becomes
\begin{equation}
H_{c}=-\dfrac{vt}{2}\sigma_{z}.\label{eq:HC-estDel}
\end{equation}
No LCH is required since $\partial_{\Delta}H$ have no level crossing.
When the OCH is applied and the system is initially prepared in a
state $\ket{\Psi}_{c\Delta}=\dfrac{1}{\sqrt{2}}\left[\ket{+x}+e^{i\beta}\ket{-x}\right]$,
the system will evolve under the Hamiltonian $H+H_{c}=\frac{\Delta}{2}\sigma_{x}$,
the state at any time $t$ is 
\begin{equation}
\ket{\psi(t)}=\dfrac{1}{\sqrt{2}}\left\{ \ket{+x}+e^{i\left[\Delta\left(t+T\right)+\beta\right]}\ket{-x}\right\} .\label{eq:Psi-t-ctrl}
\end{equation}
As one can verify, by substituting Eq. \eqref{eq:Psi-t-ctrl} into
Eq. \eqref{eq:Ig-def}, one can also obtain the expression for QFI
same as in the main text. According to Eqs. (\ref{eq:Lg-1}, \ref{eq:pm}),
the optimal measurements are the projectors formed by following vectors
\begin{equation}
\ket{\pm}_{c\Delta}=\frac{1}{\sqrt{2}}\left\{ \ket{+x}\pm ie^{i\left[\Delta\left(t+T\right)+\beta\right]}\ket{-x}\right\} .\label{eq:alphat-Del}
\end{equation}

For the estimation of $v$, taking $f_{\ket{0}}=\dfrac{v_{c}t}{2}$
and $f_{\ket{1}}=-\dfrac{v_{c}t}{2}$, the OCH becomes 
\begin{equation}
H_{c}=-\dfrac{\Delta}{2}\sigma_{x}.\label{eq:HC-v}
\end{equation}
Since the maximum and minimum eigenstates of $\partial_{v}H$ have
a level crossing at $t=0$, an OLCH $H_{LC}$ is required to avoid
the level crossing of $\partial_{v}H$ at $t=0$. According to Eqs.
(\ref{eq:Ha-general}-\ref{eq:thetak}), the OLCH is

\[
H_{LC}(t)=h(t)\left\{ \exp\left[-i\dfrac{v_{c}\left(t^{2}-T^{2}\right)}{2}\right]\ket{1}\bra{0}\right.\left.+\exp\left[i\dfrac{v_{c}\left(t^{2}-T^{2}\right)}{2}\right]\ket{0}\bra{1}\right\} ,
\]
which can also be written alternatively
\begin{equation}
H_{LC}=h(t)\left\{ \cos\left[\dfrac{v_{c}\left(t^{2}-T^{2}\right)}{2}\right]\sigma_{x}\right.\left.+\sin\left[\dfrac{v_{c}\left(t^{2}-T^{2}\right)}{2}\right]\sigma_{y}\right\} ,\label{eq:Ha-2}
\end{equation}
where $h(t)$ satisfies Eq. \eqref{eq:ht} with $t^{\prime}=0$. 

In practice, if one needs to estimate $v$, the value of $\Delta$
should be known to him. Thus OCH is also known to him. However, $v$
is unknown and therefore one has to choose a value $v_{c}$ before
performing an estimation. In general if $g_{c}\neq g$ the level crossing
Hamiltonian $H_{LC}$ may not be optimal hence the upper bound of
QFI may not be achieved. But in this particular example, we will show
in what follows that if $v_{c}$ is chosen arbitrarily, the corresponding
$H_{LC}$ is always optimal which can successfully keep the maximum
and minimum eigenstates of $\partial_{v}H$ from level crossing at
$t=0$, as long as $h(t)$ satisfy Eq. \eqref{eq:ht}. Thus an arbitrarily
chosen $v_{c}$ will give rise to the upper bound of the QFI. Assume
$\delta t$ is small enough such that the intensity of $H_{LC}$ is
much larger than $H$ and $H_{c}$ during the time interval $(-\delta t,\delta t)$.
Therefore the system's dynamics during this time interval is only
governed by $H_{LC}$. Suppose that at $t=-\delta t$ the system is
in state $\ket{1}$, then at $t=\delta t$, the system's state is
\begin{equation}
\exp\left[-i\intop_{-\delta t}^{\delta t}H_{LC}(t^{\prime})dt^{\prime}\right]\ket{1}=e^{-i(l+1/2)\pi\bm{n}\cdot\bm{\sigma}}\ket{0},\label{eq:Ha-evol}
\end{equation}
where 
\begin{equation}
\bm{n}=\left[\cos\left(\dfrac{v_{c}T^{2}}{2}\right),\:-\sin\left(\dfrac{v_{c}T^{2}}{2}\right),\:0\right],
\end{equation}
For $\bm{n}=\left[\cos\phi,\:\sin\phi,\:0\right]$, it is easy to
obtain
\begin{equation}
e^{-i(l+1/2)\pi\bm{n}\cdot\bm{\sigma}}\ket{1}=(-1)^{l+1}ie^{-i\phi}\ket{0},\label{eq:EulerF1}
\end{equation}
\begin{equation}
e^{-i(l+1/2)\pi\bm{n}\cdot\bm{\sigma}}\ket{0}=(-1)^{l+1}ie^{i\phi}\ket{1}.\label{eq:EulerF2}
\end{equation}
Thus, one can find the result of Eq. \eqref{eq:Ha-evol} is equal
to $\ket{0}$ up to a phase which depends on $v_{c}$, i.e., 
\begin{equation}
\exp\left[-i\intop_{-\delta t}^{\delta t}H_{LC}(t^{\prime})dt^{\prime}\right]\ket{1}=\left(-1\right)^{l+1}ie^{i\frac{v_{c}T^{2}}{2}}\ket{0}.\label{eq:Ha-on-0}
\end{equation}
Similarly,

\begin{equation}
\exp\left[-i\intop_{-\delta t}^{\delta t}H_{LC}(t^{\prime})dt^{\prime}\right]\ket{0}=\left(-1\right)^{l+1}ie^{-i\frac{v_{c}T^{2}}{2}}\ket{1}.\label{eq:Ha-on-1}
\end{equation}
An alternative geometric interpretation is that from right hand side
of Eq. \eqref{eq:Ha-evol} we find that the operation of $e^{-i(l+1/2)\pi\bm{n}\cdot\bm{\sigma}}$
on $\ket{1}$ is equivalent as rotating the state $\ket{1}$ by $(2l+1)\pi$
on the Bloch sphere. The rotation will gives us a state $\ket{0}$
up to a phase that depends on $v_{c}$. 

For $t<0$, the system evolves from initial state under the Hamiltonian
$H+H_{c}$, which is $\frac{vt}{2}\sigma_{z}$ for the optimal control
case. Thus the state is 
\begin{equation}
\ket{\psi(t)}=\dfrac{1}{\sqrt{2}}\left\{ \ket{0}+e^{i\left[\frac{v}{2}\left(t^{2}-T^{2}\right)+\beta\right]}\ket{1}\right\} .\label{eq:Psi-tleq0}
\end{equation}
Thus, according to Eqs. (\ref{eq:Lg-1}, \ref{eq:pm}), the optimal
measurements are projectors formed by the following vectors
\begin{equation}
\ket{\pm}_{cv,<}=\frac{1}{\sqrt{2}}\left\{ \ket{0}\pm ie^{i\left[\frac{v}{2}\left(t^{2}-T^{2}\right)+\beta\right]}\ket{1}\right\} .\label{eq:alpha1t-eps}
\end{equation}
For $t>0$, the calculation of $\ket{\psi(t)}$ is followed by two
steps. First, we first consider the dynamics from $t=0^{-}$ to $t=0^{+}$.
Form Eq. \ref{eq:Psi-tleq0}, we have 
\begin{equation}
\ket{\psi(0^{-})}=\dfrac{1}{\sqrt{2}}\left\{ \ket{0}+e^{i\left(-\frac{v}{2}T^{2}+\beta\right)}\ket{1}\right\} .\label{eq:Psi0-}
\end{equation}
According to Eqs. (\ref{eq:Ha-on-0}, \ref{eq:Ha-on-1}), we obtain

\begin{equation}
\ket{\psi(0^{+})}=\dfrac{1}{\sqrt{2}}\left\{ \ket{0}+e^{i\left(\frac{v}{2}T^{2}-\beta-v_{c}T^{2}\right)}\ket{1}\right\} ,\label{eq:Psi0+}
\end{equation}
where the redundant overall phase in Eq. \eqref{eq:Psi0+} is dropped.
Then we consider the dynamics from $0^{+}$ to $t$, the final state
is 
\begin{equation}
\ket{\psi(t)}=\dfrac{1}{\sqrt{2}}\left\{ \ket{0}+e^{i\left[\frac{v}{2}\left(t^{2}+T^{2}\right)-\beta-v_{c}T^{2}\right]}\ket{1}\right\} .\label{eq:Psi-tgeq0}
\end{equation}
According to Eqs. (\ref{eq:Lg-1}, \ref{eq:pm}), the optimal measurements
are the projectors formed by following vectors
\begin{equation}
\ket{\pm}_{cv,>}=\frac{1}{\sqrt{2}}\left\{ \ket{0}\pm ie^{i\left[\frac{v}{2}\left(t^{2}+T^{2}\right)-\beta-v_{c}T^{2}\right]}\ket{1}\right\} .\label{eq:alpha2t-eps}
\end{equation}

\subsection{Periodic LZ transitions with optimal controls and measurements}

The Hamiltonian to implement the periodic LZ transitions discussed
in the main text is 
\begin{equation}
H=\dfrac{1}{2}\left[\epsilon_{0}+A\cos(\omega t)\right]\sigma_{z}+\dfrac{\Delta}{2}\sigma_{x}.
\end{equation}
 For estimation of $\omega$, choosing $f_{\ket{0}}=\frac{A\cos(\omega_{c}t)}{2}$
and $f_{\ket{1}}=-\frac{A\cos(\omega_{c}t)}{2}$ , the OCH is 
\begin{equation}
H_{c}=-\dfrac{\epsilon_{0}}{2}\sigma_{z}-\dfrac{\Delta}{2}\sigma_{x}.
\end{equation}
Let us assume the system starts to evolve at $t=0$ and is measured
at $t=T=\frac{N\pi}{\omega_{c}}$. Since the maximum and minimum eigenstates
of $\partial_{\omega}H$ have a level crossing at time $\frac{n\pi}{\omega}$,
OLCHs are necessary at time points $\frac{n\pi}{\omega_{c}}(n=1,2\cdots N)$(for
the optimal case we have $\omega_{c}=\omega$). According to Eqs.
(\ref{eq:Ha-general}-\ref{eq:thetak}), we have
\begin{equation}
\theta_{\ket{0}}=-\theta_{\ket{1}}=\intop_{0}^{n\pi/\omega_{c}}\frac{A\cos(\omega_{c}t^{\prime})}{2}dt^{\prime}=0,
\end{equation}

\begin{equation}
H_{LC}(t)=h(t)\left\{ \ket{0}\bra{1}+\ket{1}\bra{0}\right\} =h(t)\sigma_{x},\label{eq:HaOmg}
\end{equation}
where 
\begin{equation}
h(t)\equiv(l+\frac{1}{2})\pi\sum_{n=1}^{N}\delta\left(t-\frac{n\pi}{\omega_{c}}\right).
\end{equation}
In addition, we need to prepare the initial state $\ket{\Psi}_{c\omega}=\frac{1}{\sqrt{2}}\left(\ket{0}+e^{i\beta}\ket{1}\right)$.
Thus the state $\ket{0}$ at $t=\left(\frac{n\pi}{\omega_{c}}\right)^{+}$
evolves to $e^{-i\phi_{n}/2}\ket{0}$ at $t=\left(\frac{(n+1)\pi}{\omega_{c}}\right)^{-}$
$(n=0,\,1,\,\cdots N)$, with no level crossing Hamiltonians applied
during this period, where
\begin{equation}
\phi_{n}\equiv\intop_{n\pi/\omega_{c}}^{(n+1)\pi/\omega_{c}}A\cos\left(\omega t\right)dt=\dfrac{A}{\omega}\left\{ \sin\left[\dfrac{(n+1)\pi\omega}{\omega_{c}}\right]-\sin\left[\dfrac{n\pi\omega}{\omega_{c}}\right]\right\} \label{eq:phi-n}
\end{equation}
is the accumulated phase during this time. Note that $\phi_{n}$'s
should be distinguished from from the notations $\phi_{0}$, $\phi$
in Sect. \ref{sec:states} and \ref{sec:Exact-QFIs}. Schematically,

\begin{equation}
\begin{array}{ccc}
t=\left(\frac{n\pi}{\omega_{c}}\right)^{+} &  & t=\left(\frac{(n+1)\pi}{\omega_{c}}\right)^{-}\\
\ket{0} & \xRightarrow[H+H_{c}=A/2\cos(\omega t)\sigma_{z}]{\text{evolve under}} & e^{-i\phi_{n}/2}\ket{0}\\
\ket{1} & \xRightarrow[H+H_{c}=A/2\cos(\omega t)\sigma_{z}]{\text{evolve under}} & e^{i\phi_{n}/2}\ket{1}
\end{array},
\end{equation}
where $n=0,\,1,\,2\cdots N$. If there were no OLCHs at $t=t_{n}$,
we would end up with at state 
\begin{equation}
\ket{\psi(T)}=\frac{1}{\sqrt{2}}\left[\ket{0}+\exp\left(i\sum_{n=0}^{N-1}\phi_{n}+i\beta\right)\ket{1}\right],
\end{equation}
which yields same QFI as one can calculate from $\left(A\int_{0}^{N\pi/\omega_{c}}t\sin(\omega t)dt\right)^{2}$.

We will see in what follows that by applying the OLCHs at $t=\left(\frac{n\pi}{\omega_{c}}\right)(n=1,2\cdots N)$,
the QFI will be dramatically improved. Using Eqs. (\ref{eq:EulerF1},
\ref{eq:EulerF2}), we have 
\begin{equation}
\exp\left[-i\intop_{n\pi/\omega_{c}-\delta t}^{n\pi/\omega_{c}+\delta t}H_{LC}(t^{\prime})dt^{\prime}\right]\ket{1}=e^{i(l+\frac{3}{2})\pi}\ket{0},
\end{equation}

\begin{equation}
\exp\left[-i\intop_{n\pi/\omega_{c}-\delta t}^{n\pi/\omega_{c}+\delta t}H_{LC}(t^{\prime})dt^{\prime}\right]\ket{0}=e^{i(l+\frac{3}{2})\pi}i\ket{1}.
\end{equation}
Thus when OLCHs are applied, we have 
\[
\ket{\psi(T)}=\frac{1}{\sqrt{2}}\left(\ket{0}+e^{i\Phi_{N}}\ket{1}\right),
\]
where 
\begin{equation}
\Phi_{N}=\sum_{n=0}^{N-1}(-1)^{n}\phi_{n}+(-1)^{N}\beta+N(l+\frac{3}{2})\pi,
\end{equation}
which yields QFI $\left[\sum_{n=0}^{N-1}(-1)^{n}\partial_{\omega}\phi_{n}\right]^{2}$.
Substituting of Eq. \eqref{eq:phi-n} into this expression for QFI
and then set $\omega_{c}=\omega$ one obtains the same result as in
the main text. Since $\omega_{c}=\omega$, $\phi_{n}=0$ for the optimal
case where the first term in $\Phi_{N}$ vanishes, according to Eqs.
(\ref{eq:Lg-1}, \ref{eq:pm}), the optimal measurements are the projectors
formed by following vectors

\begin{equation}
\ket{\pm}_{c\omega}=\dfrac{1}{\sqrt{2}}\left\{ \ket{0}\pm i^{N+1}(-1)^{Nl+N}\exp\left[i(-1)^{N}\beta\right]\ket{1}\right\} .
\end{equation}
If the estimation time is at $T=\frac{(N+\alpha)\pi}{\omega_{c}}$,
aside from the phase accumulation from $t=0$ to $t=\frac{N\pi}{\omega}$,
the phase also accumulate from $t=\frac{N\pi}{\omega}$ to $t=\frac{(N+\alpha)\pi}{\omega_{c}}$,
the state at $t=T$ becomes
\[
\ket{\psi(T)}=\frac{1}{\sqrt{2}}\left\{ \ket{0}+\exp\left[i\Phi_{N}+i(-1)^{N}\phi_{N+1-\alpha}\right]\ket{1}\right\} ,
\]
where 
\begin{equation}
\phi_{N+1-\alpha}\equiv\intop_{N\pi/\omega_{c}}^{(N+\alpha)\pi/\omega_{c}}A\cos\left(\omega t\right)dt=\dfrac{A}{\omega}\left\{ \sin\left[\dfrac{(N+\alpha)\pi\omega}{\omega_{c}}\right]-\sin\left[\dfrac{N\pi\omega}{\omega_{c}}\right]\right\} .\label{eq:phi-alpha}
\end{equation}
Setting $\omega_{c}=\omega$, the optimal measurements are the projectors
formed by following vectors: 

\begin{eqnarray}
\ket{\pm}_{c\omega} & = & \dfrac{1}{\sqrt{2}}\left\{ \ket{0}\pm i^{N+1}(-1)^{Nl+N}\exp\left[i(-1)^{N}\dfrac{A\sin\left(N\pi+\alpha\pi\right)}{\omega}+i(-1)^{N}\beta\right]\ket{1}\right\} \\
 & = & \dfrac{1}{\sqrt{2}}\left\{ \ket{0}\pm i^{N+1}(-1)^{Nl+N}\exp\left[i\dfrac{A\sin\left(\alpha\pi\right)}{\omega}+i(-1)^{N}\beta\right]\ket{1}\right\} .
\end{eqnarray}

\section*{Appendix A: Digamma functions \label{sec:Digamma-functions}}

The integral representation for the digamma function $\varphi(z)$
is \cite{whittaker_course_1996,wang_special_1989}

\begin{equation}
\varphi(z)\equiv\left[\ln\Gamma(z)\right]^{\prime}=\intop_{0}^{\infty}\left\{ \dfrac{e^{-t}}{t}-\dfrac{e^{-\Re(z)t}e^{-i\Im(z)t}}{1-e^{-t}}\right\} dt=\intop_{0}^{\infty}\left\{ \dfrac{e^{-t}}{t}-\dfrac{e^{-\Re(z)t}}{1-e^{-t}}\cos\left[\Im(z)t\right]\right\} dt+i\intop_{0}^{\infty}\left\{ \dfrac{e^{-\Re(z)t}}{1-e^{-t}}\sin\left[\Im(z)t\right]\right\} dt.\label{eq:digamma}
\end{equation}
Upon defining the line integrals
\begin{equation}
\theta_{\nu}(a)\equiv\intop_{0}^{\infty}\left\{ \dfrac{e^{-t}}{t}-\dfrac{e^{-\nu t}}{1-e^{-t}}\cos\left(at\right)\right\} dt,\label{eq:theta-nu}
\end{equation}

\begin{equation}
\eta_{\nu}(a)\equiv\intop_{0}^{\infty}\left\{ \dfrac{e^{-\nu t}}{1-e^{-t}}\sin\left(at\right)\right\} dt,
\end{equation}
we may write $\varphi\left(1-ia\right)$ as
\begin{equation}
\varphi\left(1-ia\right)=\theta_{1}\left(a\right)-i\eta_{1}\left(a\right).\label{eq:varphi1}
\end{equation}
Note that $\eta_{1}(a)$ can be computed in elementary functions (\cite[3.951.12 in pp496]{zwillinger_table_2014}):

\begin{equation}
\eta_{1}(a)=\intop_{0}^{\infty}\dfrac{e^{-t}\sin(at)}{1-e^{-t}}dt=\intop_{0}^{\infty}\dfrac{\sin(at)}{e^{t}-1}dt=\dfrac{\pi a\coth\left(\pi a\right)-1}{2a}.\label{eq:eta1_closed}
\end{equation}

\end{document}